\documentclass{aa}  

\usepackage[varg]{txfonts}
\usepackage{amssymb}
\usepackage{graphicx}

\usepackage{graphicx}	
\usepackage{mathtools}	
\usepackage{amssymb}	
\usepackage{multicol}        
\usepackage{bm}		
\usepackage{pdflscape}	
\usepackage{xcolor}
\usepackage{placeins}
\usepackage{txfonts}
\begin{document} 

   \title{The backreaction of stellar wobbling on accretion discs of massive protostars}


   \author{D. M.-A.~Meyer
          \inst{1}
          \and
          E. Vorobyov\inst{2,3}
          }

    \institute{Institute of Space Sciences (ICE, CSIC), Campus UAB, Carrer de Can Magrans s/n, 
    08193 Barcelona, Spain\\
    \email{dmameyer.astro@gmail.com}
    \and
     Ural Federal University, 19 Mira Str., 620002 Ekaterinburg, Russia
    \and
     Research Institute of Physics, Southern Federal University, Rostov-on-Don 344090, Russia  \\
          }
   \date{}

 
  \abstract
   {
In recent years, it has been demonstrated that  massive stars see their infant circumstellar medium shaped into a large, irradiated, gravitationally 
unstable accretion disc during their early formation phase.
%
%
Such discs constitute the gas reservoir in which nascent high-mass stars gain substantial 
fraction of their mass by episodic 
accretion of dense gaseous circumstellar clumps, simultaneously undergoing accretion-driven bursts 
and producing close-orbit spectroscopic companions to the young high-mass stellar object.  
   }
   {
    We aim to evaluate the effects of stellar motion, caused by the disc non-axisymmetric gravitational field, on the disc evolution and its spatial morphology. In particular, we analyze the disc propensity to gravitational instability and fragmentation, and also disc appearance on synthetic millimeter-band images pertinent to the {\sc alma} facility.
   }
   {
We employed three-dimensional radiation-hydrodynamical simulations of the 
surroundings of a young massive star in the non-inertial spherical coordinate system, adopting the highest spatial resolution to date and including the indirect star-disc gravitational 
potential caused by the asymmetries in the circumstellar disc. The resulting disc configurations were postprocessed with the radiation transfer tool {\sc RADMC-3D} and {\sc CASA} softwear to obtain disc synthetic images.
   }
   {
   We confirm that the early evolution of the accretion disc is notably different when stellar 
wobbling is taken into account. The redistribution of angular momentum in the system makes the 
disc smaller and rounder, reduces the number of circumstellar gaseous clumps formed via disc gravitational fragmentation, and 
prevents the ejection of gaseous clumps from the disc. 
The synthetic predictive images at millimeter wavelengths of the accretion disc including 
stellar wobbling are in better agreement with the observations of the surroundings of massive 
young stellar objects, namely, \textcolor{black}{
AFGL 4176 mm1, G17.64+0.16 and G353.273},   than our numerical hydrodynamics simulations omitting this physical mechanism.    
   }
{
Our work confirms that stellar wobbling is an essential ingredient to account for in numerical simulations of accretion discs of massive protostars. 
   }


   \maketitle
%


\section{Introduction}

Although the formation of massive stars is a rare event, those objects are preponderant 
engines in the cycle of life of matter in galaxies. The mystery of their birth principally 
lies in the high optical depth of the parent pre-stellar cores in which they form. 
Apart from revealing their common multiplicity~\citep{kraus_apj_835_2017}, the quest 
for the first observation of Keplerian-orbiting material around massive young stellar 
objects (MYSOs) saw a first breakthrough with the works of~\citet{johnston_apj_813_2015,
ilee_mnras_462_2016,forgan_mnras_463_2016,2018arXiv180410622G,maud_aa_620_2018}.  
The study of MYSOs continued with the hunt for evidence of filamentary/clumpy substructures
in discs around W33A MM1-Main~\citep{maud_467_mnras_2017} 
G350.69-0.49~\citep{chen_apj_835_2017} and G11.92-0.61 MM1~\citep{2018ApJ...869L..24I}. 
Besides, accretion variability onto high-mass protostars has been monitored in 
massive young stellar objects~\citep{keto_apj_637_2006,stecklum_2017a} and in pulsed bipolar
outflows~\citep{Cunningham_apj_692_2009,cesaroni_aa_509_2010,caratti_aa_573_2015,
purser_mnras_460_2016,reiter_mnras_470_2017,burns_mnras_467_2017,burns_iaus_336_2018,
purser_mnras_475_2018,samal_mnras_477_2018}.

Since the variability of the accretion flow onto the surface of massive protostars 
governs their pre-main-sequence evolution in the 
Hertzsprung-Russell diagram~\citep{meyer_mnras_484_2019}, the 
study of massive young stellar objects is therefore tied to the understanding of 
the inhomogeneities and substructures in their circumstellar discs as likely causes of the accretion variability.
These non-axisymmetric disc features are
possibly produced 
by instabilities of self-gravitating nature, which were extensively studied in 
the context of low-mass stellar objects~\citep{papaloizou_mnras_248_1991,
pickett_apj_590_2003,2005ApJ...621L..69R,rafikov_apj_662_2007}. 
Several questions remain wide open, such as MYSOs chemical 
evolution~\citep{ahmadi_aa_618_2018,2023arXiv231207184G} and the role of magnetic fields in 
the regulation of their fragmentation~\citep{hennebelle_apj_830_2016}.

The circumstellar medium of massive protostars 
has been investigated using computationally-intensive numerical 
simulations, see e.g.~\citet{krumholz_apj_665_2007,krumholz_apj_656_2007,
krumholz_sci_323_2009,harries_mnras_448_2015,harries_2017}. 
Such calculations raised a number of questions and revealed technical challenges 
caused by the three-dimensional intrinsic nature of the problem and by the multiple 
physical processes involved in the formation mechanism of massive stars 
(cooling and heating of the gas, protostellar irradiation and radiation 
transport into the disc, disc wind, non-ideal magneto-hydrodynamical processes, 
etc.). 
Amongst many approaches developed to tackle the problem of massive star 
formation, the new paradigm known as the burst mode of accretion produced results consistent 
with not only models and observations of the lower mass regimes of star formation, 
but also with direct observations of forming massive stars.

The burst mode of accretion is a depiction of star-forming processes which 
fills the gap between the episodic accretion onto protostars and the 
luminous bursts originating from young stellar objects and observed in some 
star-forming regions~\citep{vorobyov_apj_633_2005,voroboyov_apj_650_2006, 
vorobyov_apj_719_2010,Machida_2011,vorobyov_apj_805_2015}. 
In this picture, 
clumpy circumstellar material from gravitationally-unstable accretion disc around young 
stars resulting from the infall of the molecular material of pre-stellar cores 
inward-migrate from the disc to the protostellar surface, hence triggering a 
sudden luminosity rise. This provides an explanation to the so-called 
FU-Orionis accretion-driven burst phenomenon and offers evident connections to 
the pre-zero-age-main-sequence stellar evolution~\citep{elbakyan_mnras_484_2019} and the 
prediction/observations of substructures in discs~\citep{dong_apj_823_2016}. 
The general idea of the burst mode of accretion was extended and elaborated by 
Nayakshin and co-authors to better explain the burst phenomenon and include 
planet formation in inward-migrating clumps~\citep{2010MNRAS.408L..36N,nayakshin_mnras_426_2012,
nayakshin_mnras_461_2016,2017PASA...34....2N,2018A&A...618A...7V,elbakyan_aa_651_2021,meyer_mnras_518_2023}.

Interestingly, this picture originally developed for the low-mass regime of star formation 
can be naturally extended to the primordial and massive regimes of star formation.  
In~\citet{meyer_mnras_464_2017} and~\citet{meyer_mnras_473_2018} it was shown that a 
self-consistent picture unifying disc fragmentation and binary formation in the context 
of massive young stellar objects is possible, linking both accretion-driven outbursts 
as monitored in the MYSOs S255IR-NIRS3 and NGC 6334I-MM1 to the close/spectroscopic 
companions of some massive stars~\citep{2013A&A...550A..27M,2014ApJS..213...34K,chini_424_mnras_2012}. 
Particularly, accretion-driven outbursts are interpreted as to be the observational 
signature of the presence of a self-gravitating disc shaped by efficient 
gravitational instability. The bursts occur in series of multiple eruptions and 
their intensity is a function of the mass of the accreted 
material~\citep{meyer_mnras_482_2019}. Such episodic flares witness the sudden 
increment of the protostellar mass with dense, compact circumstellar 
material~\citep{meyer_mnras_518_2023}  that affect the internal structure 
of the protostars, profoundly changing their surface properties and 
provoking excursions in the Hertzsprung-Russel diagram~\citep{meyer_mnras_484_2019}. 
%


\begin{table*}
	\centering
	\caption{
	List of the simulation models performed in our study. 
	The table provides the initial mass of the molecular pre-stellar core 
    $M_{\rm c}$ (in $M_{\odot}$), its initial rotational-to-gravitational 
    energy ratio $\beta$ (in $\%$) and the final simulation time $t_{\rm end}$ 
    in each model. The last column indicates whether the simulation 
    includes stellar wobbling or not. 
	}
	\begin{tabular}{lccccccr}
	\hline
	$\mathrm{Models}$  &  $M_{\rm c}$ $(M_{\odot})$  &  $\mathrm{Grid}\, \mathrm{resolution}$   &  $\beta$ ($\%$)  &  $t_{\rm end}$ ($\rm kyr$)        & $\mathrm{Wobbling}$ \\ 
	\hline   
	{\rm Run-512-100$\rm M_{\odot}$-4$\%$-wio}  &  $100$  &  $512\times81\times512$  & $4$  & 32.65 & no  \\  	
	{\rm Run-512-100$\rm M_{\odot}$-4$\%$-wi}   &  $100$  &  $512\times81\times512$  & $4$  & 32.65 & yes \\ 
        \hline
	\end{tabular}
\label{tab:models}\\
\end{table*}

Numerical results have preceded the direct observation of an accretion disc around 
a young high-mass star. Such predictions divide into two distinct categories of 
models, obtained either using a Cartesian grid with an adaptive mesh refinement method 
and a protostar allowed to move freely inside the computational 
domain~\citep{krumholz_apj_665_2007,Krumholz_sci_2009,peters_apj_711_2010,
commercon_apj_742_2011,
seifried_mnras_417_2011,seifried_mnras_422_2012,seifried_mnras_432_2013,
klassen_apj_797_2014,seifried_mnras_446_2015,klassen_apj_823_2016,
rosen_mnras_463_2016,mignon_risse_aa_652_2021,commercon_aa_658_2022,mignonrisse_673_aa_2023}, 
or using a spherical grid with a static logarithmic mesh and assuming that the 
growing protostar is fixed to the origin of domain~\citep{kuiper_apj_722_2010,
kuiper_aa_511_2010,kuiper_apj_732_2011,kuiper_apj_763_2013,kuiper_apj_772_2013,
meyer_mnras_464_2017,meyer_mnras_473_2018,meyer_mnras_482_2019,
meyer_487_MNRAS_2019,ahmadi_aa_632_2019,oliva_aa_644_2020,meyer_mnras_500_2021, 
oliva_aa_669_2023,oliva_aa_669_II_2023}. 
In both cases, the initial conditions and the included microphysical processes 
in the models are similar (self-gravity, radiation transport coupled to 
proto-stellar evolution).

Cartesian grids have the advantage not to have coordinate singularities and to permit a 
straightforward implementation of the stellar motion, however, the inclusion 
of stellar radiation transfer into the models makes calculations computationally expensive. 
Spherical grids allow us to reach high spatial resolution in the inner circumstellar environment of 
protostars, which precisely permitted to reveal the burst mode of accretion 
in the context of massive star formation~\citep{meyer_mnras_464_2017}. 
Both computing philosophies have been thoroughly compared on the basis of high-spatial 
resolution simulations in~\citet{Mignon_aa_672_2023} and the authors conclude 
therein in a qualitative agreement between both approaches, the remaining differences 
in the results being minor.

Nevertheless, we showed in~\citet{meyer_mnras_517_2022} 
that when permitting the protostar to move, as a consequence of the disc-star 
gravitational interaction, 
stellar motion delays disc fragmentation and affects the millimeter continuum 
emission properties~\citep{meyer_mnras_517_2022}. 
Such an effect is not encompassed in the comparison study of~\citet{Mignon_aa_672_2023}. 
Difference between the spherical models of ~\citet{meyer_mnras_517_2022} and the 
Cartesian simulations of ~\citet{Mignon_aa_672_2023} lie in the spatial resolution 
and the absence/presence of stellar wobbling. It is hereby proposed to perform 
a comparison of disc simulation with the resolution of~\citet{Mignon_aa_672_2023}, 
with and without the inclusion of stellar inertia.

This study aims at investigating the effects of finite stellar inertia onto the formation and growth of 
circumstellar accretion discs around massive protostars. We also address the question 
of the millimeter-observability by continuum dust emission of such disc nebulae 
and of the substructures forming in them by gravitational instability. 
To this end, a pair of three-dimensional gravito-radiation-hydrodynamics numerical 
simulations is performed, resolved with the highest to-date spatial resolution used in 
spherically-based calculation in the study of~\citet{Mignon_aa_672_2023}. Two scenarios 
are considered: without and with stellar inertia, namely, without and with the star motion in response to the gravitational 
interaction with the disc. 
Selected characteristic outcomes of the simulations are post-processed with radiation 
transfer and imaging methods in order to produce synthetic $1.3\, \rm mm$ dust 
continuum images for the {\it Atacama Large Millimeter/submillimeter 
Array} ({\sc alma}) interferometer operating in its Band 6, see the work 
of~\citet{meyer_mnras_473_2018} and ~\citet{jankovic_mnras_482_2019}. 
We explore in these images the indirect effects of stellar inertia onto the disc 
structure, its eventual fragmentation by efficient gravitational instability, 
the formation of gaseous clumps therein, as well as their dynamics inside 
(migration) and outside (ejection) of the disc. 
Finally, the most realistic high-resolution disc model, possessing both a high-spatial 
resolution and stellar inertia included in it, is compared to observations available 
in the literature.

This work is organized as follows. In Section~\ref{sect:method}, the methods 
used to perform high-resolution hydrodynamical simulations of a realistic 
accretion disc around a young massive star are presented, together with the 
radiation transfer procedure utilised to predict its continuum thermal dust 
emission at the milliliter waveband. 
Section~\ref{sect:results}, introduce the reader to the results
and details of the hydrodynamical simulations of accretion discs 
surrounding massive protostars. 
The radiative transfer calculations for dust continuum emission of the stellar 
surroundings of these forming massive stars are shown in Section~\ref{sect:discussion}, 
for several characteristic time instances of the disc evolution. 
Conclusions are finally provided in Section~\ref{sect:cc}.


\section{Method}
\label{sect:method}

In this section, the methods used to simulate and predict the millimeter 
continuum appearance as seen by the {\sc ALMA} interferometre of the circumstellar 
medium of a forming massive protostar are presented.

\subsection{Radiation-hydrodynamics simulations}
\label{sect:hd}

Numerical simulations of accretion discs around protostars were carried out under the 
assumption of a midplane-symmetric computational domain. It is initialised with a 
$M_{\rm c}=100~M_{\odot}$ pre-stellar core of uniform temperature $T_{\rm c}=10\, \rm K$ 
which is rigidly rotating. The initial mass density distribution is spherically symmetric and 
profiled as, 
\begin{equation}
    \rho(r) = \frac{ 3 }{ 8\pi  } \frac{ M_{\rm c} }{  R_{\rm c}^{ 3/2 } } r^{ -3/2 },
    \label{eq:density_profile}
\end{equation}
where $r$ is the radial coordinate and $R_{\rm c}$ the outer core radius, respectively. 
The gas angular velocity of the pre-stellar core is initialised according to the 
distribution $\Omega(R) \propto R^{ -3/2 }$, with the cylindrical radius 
$R = r \sin(\theta)$, and according to the rotational-to-gravitational energy ratio,
%
%
%
\begin{equation}
  \beta =  \frac{   E_{\rm rot}   }{ E_{\rm grav}  }  = 0.04 \propto
    \frac{ 1 }{ 7 }    
	\frac{ 1 }{  G M_{\rm c}  R_{\rm c}^{ -3 }  }     
    \int_{ 0 }^{ \pi } d\theta \sin( \theta )^{ 3 }.
    \label{eq:beta2}    
\end{equation}

The grid mesh $[r_{\rm in},R_{\rm c}]\times[0,\pi/2]\times[0,2\pi]$ mapping the computational 
domain expands logarithmically along the radial direction $r$, as a cosine in the polar direction 
$\theta$ and is uniform along the azimuthal direction $\phi$. It is made of 
$N_{\rm r}=512\times\,N_{\rm \theta}=81\times\,N_{\rm \phi}=512$ grid zones, respectively. 
The inner radius $r_{\rm in}$ constitutes a semi-permeable sink cell fixed onto the origin 
of the domain and the outer radius, assigned to outflow boundary conditions, is located at 
$R_{\rm c}=0.1\, \rm pc$. This study adopts $r_{\rm in}=20\, \rm au$,  
which permits the highly-resolved models to reach long integration times $t_{\rm end}$ 
without dealing with dramatic time-step restrictions in the innermost grid zones. 
The utilised grid spatially resolves the inner region of the midplane where the disc 
fragments are accreted, while keeping the overall number of grid cells to a decent value 
and thus reducing the numerical cost of this computationally-intensive calculation.

The numerical simulations follow the gravitational collapse of the pre-stellar core and the early 
evolution of circumstellar disc around the massive protostar, accounting for both the 
stellar direct irradiation of protostellar photospheric photons and the radiation 
transport into the circumstellar medium. 
The accretion rate is calculated onto the protostar as the material loss $\dot{M}$ 
through the sink cell and the surface and intrinsic properties of the protostar, 
such as the stellar radius and the photospheric luminosity, are time-dependently 
interpolated using the pre-calculated protostellar evolutionary tracks 
of~\citet{hosokawa_apj_691_2009}, see 
also~\citet{meyer_mnras_464_2017,meyer_mnras_473_2018}.  
Two models are performed with the above described initial conditions, with and 
without the inclusion of stellar inertia (Table \ref{tab:models}).

\subsection{Governing equations}
\label{sect:equations}

The set of equations describing the dynamics of the pre-stellar core collapse reads, 
\begin{equation}
	   \frac{\partial \rho}{\partial t}  + 
	   \vec{ \nabla }  \cdot ( \rho \vec{v} )  =   0,
\label{eq:euler1}
\end{equation}
\begin{equation}
	   \frac{\partial \rho \vec{v} }{\partial t}  + 
           \vec{\nabla} \cdot ( \rho  \vec{v} \otimes \vec{v})  + 
           \vec{\nabla}p 			      =   \vec{f},
\label{eq:euler2}
\end{equation}
\begin{equation}
	  \frac{\partial E }{\partial t}   + 
	  \vec{\nabla} \cdot ((E+p) \vec{v})  =	   
	  \vec{v} \cdot \vec{f} ,
\label{eq:euler3}
\end{equation}
which are the relations for the conservation of the mass, the momentum and the energy of the 
infalling material. In the above relations, the gas density reads $\rho$, the thermal pressure 
$p=(\gamma-1)E_{\rm int}$, the gas velocity $\vec{v}$ and $\gamma=5/3$ is the adiabatic 
index. Hence,  
\begin{equation}
    E  = E_{\rm int} + \rho \frac{ \vec{v}^{2} }{ 2 }
    = \frac{ p }{ (\gamma-1) } + \rho \frac{ \vec{v}^{2} }{ 2 }, 
\end{equation}
is the total energy of the gas. 
The right-hand term of the momentum and energy conservation laws are the force density vector, 
\begin{equation}
	  \vec{ f } = -\rho \vec{\nabla} \Phi_{\rm tot} 
			- \lambda \vec{\nabla} E_{\rm R} 
			- \vec{\nabla} \cdot \Big( \frac{ \vec{F_{\star}} }{ c } \Big) \vec{e}_{\rm r},
\label{eq:f}
\end{equation}
with the flux limiter $\lambda=1/3$, the thermal radiation energy density $E_{\rm R}$, the 
radial unit vector $\vec{e}_{\rm r}$, the protostellar radiation flux $\vec{F_{\star}}$
and the speed of light $c$. The quantity $\Phi_{\rm tot}$ is the total gravitational 
potential of the gas and the star, including the indirect potential as described below. 

\begin{figure*}
        \centering
        \includegraphics[width=0.82\textwidth]{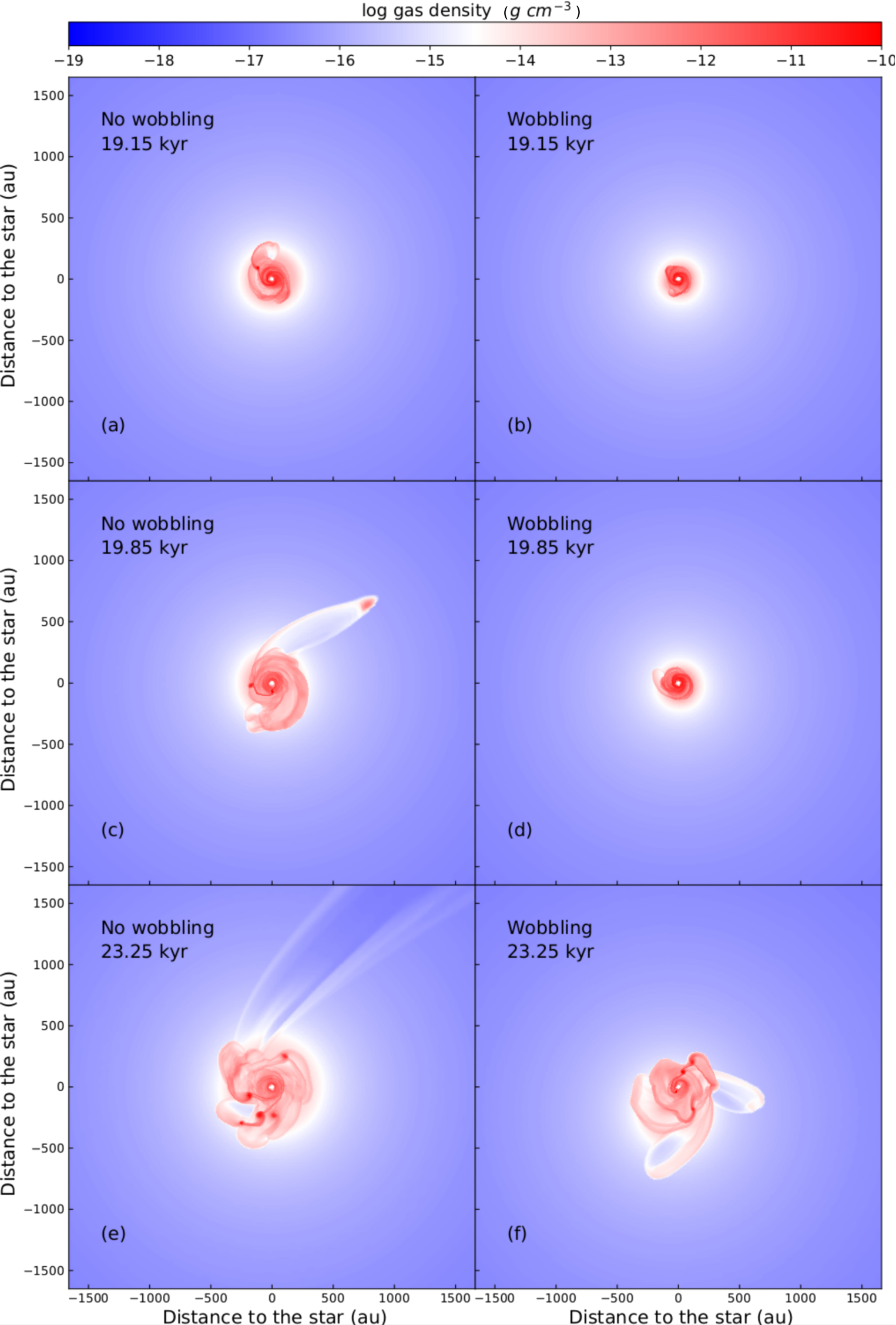}  \\
        \caption{
        Mid-plane density fields (in $\rm g\, \rm cm^{-3}$ ) in the 
        accretion disc hydrodynamical simulations performed without (left panels) 
        and with (right panels) stellar inertia. The figures are shown 
        for several time instances, namely $19.15$ (a,b), $19.85$ (c,d) and 
        $23.25\, \rm kyr$ (e,f). 
        The initial kinetic-to-gravitational energy ratio of the collapsing 
        molecular cloud is $\beta = 4\%$. 
}
        \label{fig:disc_density_1}  
\end{figure*}

\begin{figure*}
        \centering
        \includegraphics[width=0.82\textwidth]{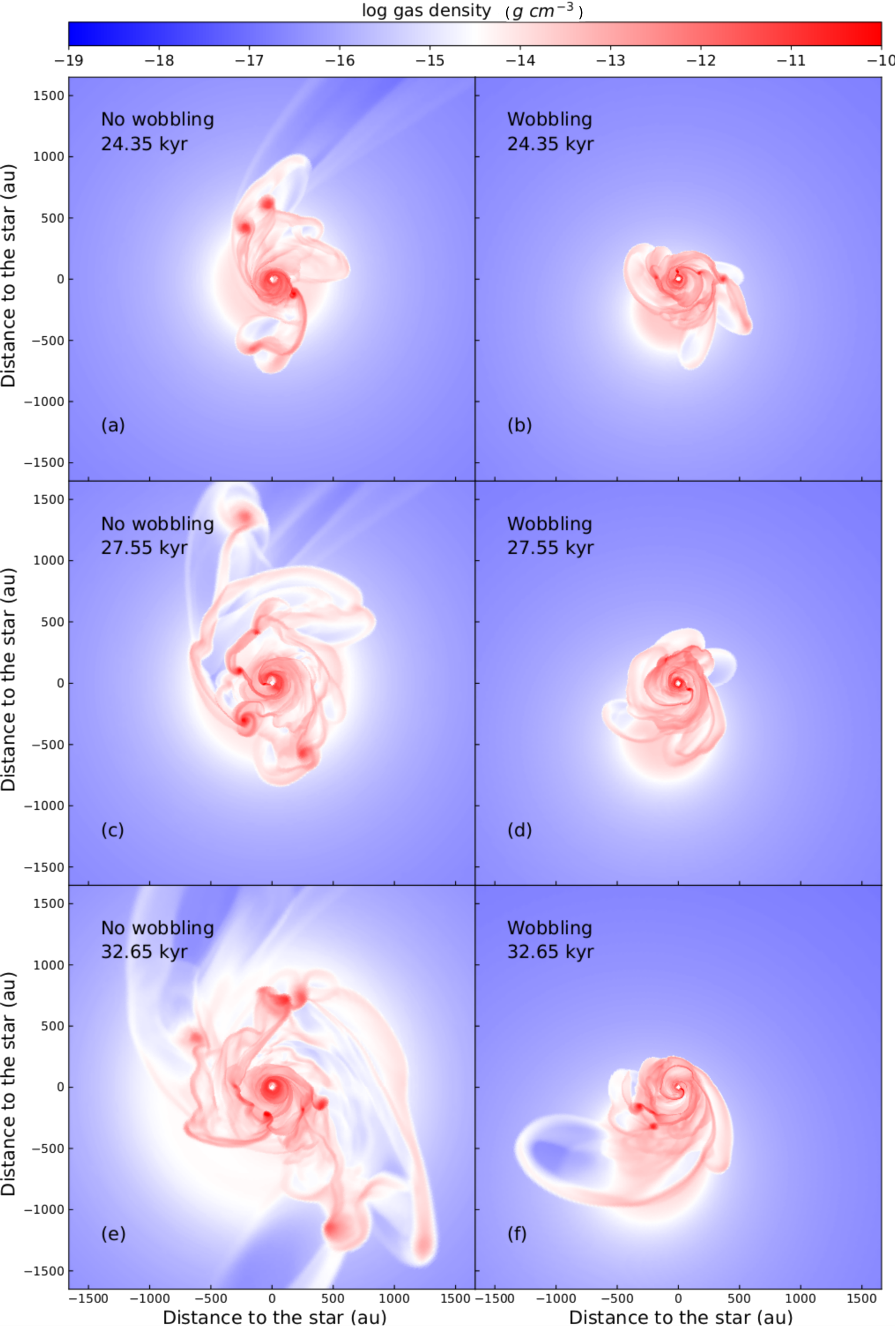}  \\
        \caption{
        As Fig. \ref{fig:disc_density_1}, for the time instances  
        $24.35$ (a,b), $27.55$ (c,d) and $32.65\, \rm kyr$ (e,f). 
}
        \label{fig:disc_density_2}  
\end{figure*}

\subsection{Stellar inertia}
\label{sect:inertia}

This numerical setup uses the implementation for stellar motion presented in 
\citet{regaly_aa_601_2017,Hirano_2017Science,meyer_mnras_517_2022}. It is 
implemented as an additional indirect potential, $\Phi_\mathrm{wobbling}$ 
together with its associated force $\vec{F}_\mathrm{wobbling}$. 
This indirect force is implemented as the following acceleration, 
\begin{equation}
	\vec{g}^{'} + \frac{ \vec{F}_\mathrm{disc/\star} }{ M_{\star}} = 0,  
	\label{eq:acceleration}
\end{equation}
where the disc-to-protostar gravitational force reads,  
\begin{equation}
	\vec{F}_\mathrm{disc/\star} = - G M_{\star} \int_\mathrm{disc}  
    \frac{\rho(r) \, dV}{r^2} \vec{e}_\mathrm{r}, 
	\label{eq:force_int_B}
\end{equation}
with the protostellar mass $M_{\star}$, the gravitational constant G, 
the radial unit vector $\vec{e}_\mathrm{r}$ and the mass in a volume 
element $d M_\mathrm{disc}(r)=\rho(r) d V$.  
%
%
%
Hence, the total gravitational potential reads $\Phi_{\rm tot}=\Phi_{\rm sg}+\Phi_\mathrm{wobbling}$, where $\Phi_{\rm sg}$ is the input from the disc, the envelope, and the star. 
The purpose of this study consists in exploring the effects of the stellar inertia onto the 
dynamics and morphology of accretion discs at a spatial resolution exceeding that of the 
previous study of this series, see \citet{meyer_mnras_517_2022}. 

\FloatBarrier
%

%
%
%
%
%

\subsection{Numerical methods}
\label{sect:numerics}

The equations of gravito-radiation-hydrodynamics are solved with the {\sc pluto} 
code\footnote{http://plutocode.ph.unito.it/}~\citep{mignone_apj_170_2007,migmone_apjs_198_2012}. 
The direct irradiation feedback of the protostar and the radiation transport in the accretion 
disc are both taken into account within the gray approximation. It uses the numerical scheme presented in~\citet{kolb_aa_559_2013}\footnote{http://www.tat.physik.uni-tuebingen.de/\,pluto/pluto\_radiation/} and adapted in~\citet{meyer_mnras_473_2018} for the study of massive star formation. 
It is an algorithm which first ray-traces photon packages from the protostellar surface to the 
disc and then diffuses their propagation into it in the flux-limited approximation. This 
approach allows us to accurately consider both the inner heating and the outer cooling of 
our irradiated accretion discs, see~\citet{vaidya_apj_742_2011}.  
We note that similar radiation-hydrodynamics methods are also presented 
in~\citet{commercon_aa_529_2011},~\citet{flock_aa_560_2013} and~\citet{bitsch_aa_564_2014}.

The opacity description as well as the estimate of the local dust properties are 
similar as in~\citet{meyer_mnras_473_2018}, where the gas $T_{\rm gas}$ and dust $T_{\rm dust}$ 
temperatures are calculated assuming the equilibrium between the silicate grains temperature 
and the total radiation field. 
Stellar gravity is modelled by calculating the total gravitational potential of the 
central protostar and include the self-gravity of the gas by solving the Poisson 
equation using the PETSC library\footnote{https://www.mcs.anl.gov/petsc/}. 
We neglect turbulent viscosity by assuming that the most efficient mechanism for the 
transport of angular momentum are the gravitational torques in a self-gravitating disc. The 
effects of the stellar inertia onto the disc wobbling in included as described 
in~\citet{meyer_mnras_482_2019}.  
The reader interested in further reading about the method can refer to the other 
papers in this series, initiated with~\citet{meyer_mnras_464_2017}.

\subsection{Construction of synthetic disc images}
\label{sect:radiative}

As a diagnostic to evaluate the effects of stellar inertia on realistic, high-resolution 
models of the circumstellar medium of massive protostars, synthetic {\sc alma} images 
of the accretion discs were calculated using the post-processing method presented in 
\citet{meyer_487_MNRAS_2019}. 
For each panel displayed in Figs. \ref{fig:disc_density_1}-\ref{fig:disc_density_2} 
the dust density fields of the embedded accretion discs simulated with the {\sc pluto} 
code were imported into the radiative transfer code 
{\sc radmc-3d}\footnote{http://www.ita.uni-heidelberg.de/dullemond/software/radmc-3d/} 
~\citep{dullemond_2012}. 
The standard dust to gas mass ratio of 0.01 was assumed when converting 
the gas density into that of dust.
Then, the dust temperature is derived by Monte-Carlo calculation on the basis of the 
dust density, using the method presented in~\citet{bjorkman_apj_554_2001} and using 
$10^{10}$ photons packages ray-traced from the protostellar surface to the outer region 
of the accretion disc. The proper radiative transfer calculation against dust opacity 
of the star light scattered into the dusty disc is subsequently performed for a 
monochromatic light of wavelength centered onto $1.2\, \rm mm$ ($249.827\, \rm Ghz$ 
with a channel width of $50.0\, \rm Mhz$), considering that the dust is a 
\citet{laor_apj_402_1993} mixture of silicate particles. 
The photopshere is modelled as a black body of effective temperature $T_{\rm eff}$ that 
is derived for each selected simulation snapshots as a function of the age of the 
protostar, i.e. its mass and current accretion rate, which are used to bilinearly 
interpolate the protostellar evolutionary tracks of~\citet{hosokawa_apj_721_2010}.

Finally, synthetic images of the accretion discs were generated using the Common Astronomy 
Software Applications {\sc casa}\footnote{https://casa.nrao.edu/}~\citep{McMullin_aspc_376_2007} using the radiation flux output from  {\sc radmc-3d}. 
No inclination angle for the accretion disc with respect to the plane of the sky  
is assumed throughout the whole simulation process and images were 
produced with a field of view of $2000\, \rm au$ around the growing protostar. 
The simulated {\sc alma} interferometric images were thus obtained. They can either serve as a 
test for the observability of the simulated accretion discs, or as a tool to directly compare the disc models with available and/or forthcoming observations to be performed 
by the {\sc alma} facility.

As in \citet{meyer_487_MNRAS_2019}, it is assumed that the conditions for the synthetic 
observations are ideal. In other words, the pattern of telescopes/antennae are considered 
to be in their most extended spatial configuration at the Llano de Chajnantor plateau, 
permitting the acquisition of long-baseline images resolved with the maximal possible 
spatial resolution with the smallest beam size. This resolution is about 
$0.015^{"}$ for the C43-10 configuration that makes use of 43 $12\, \rm m$ antennae. 
The modelling of these Cycle 10 synthetic {\sc alma} observations assumes an exposure 
time of $10$ minutes and a precipitable water vapour $\rm pwv=0.6$. 
The assumed distance to the source is taken to $1\, \rm kpc$, which corresponds to 
that of the Earth's closest high-mass star-forming region Orion.


\begin{figure*}
        \centering
        \includegraphics[width=0.82\textwidth]{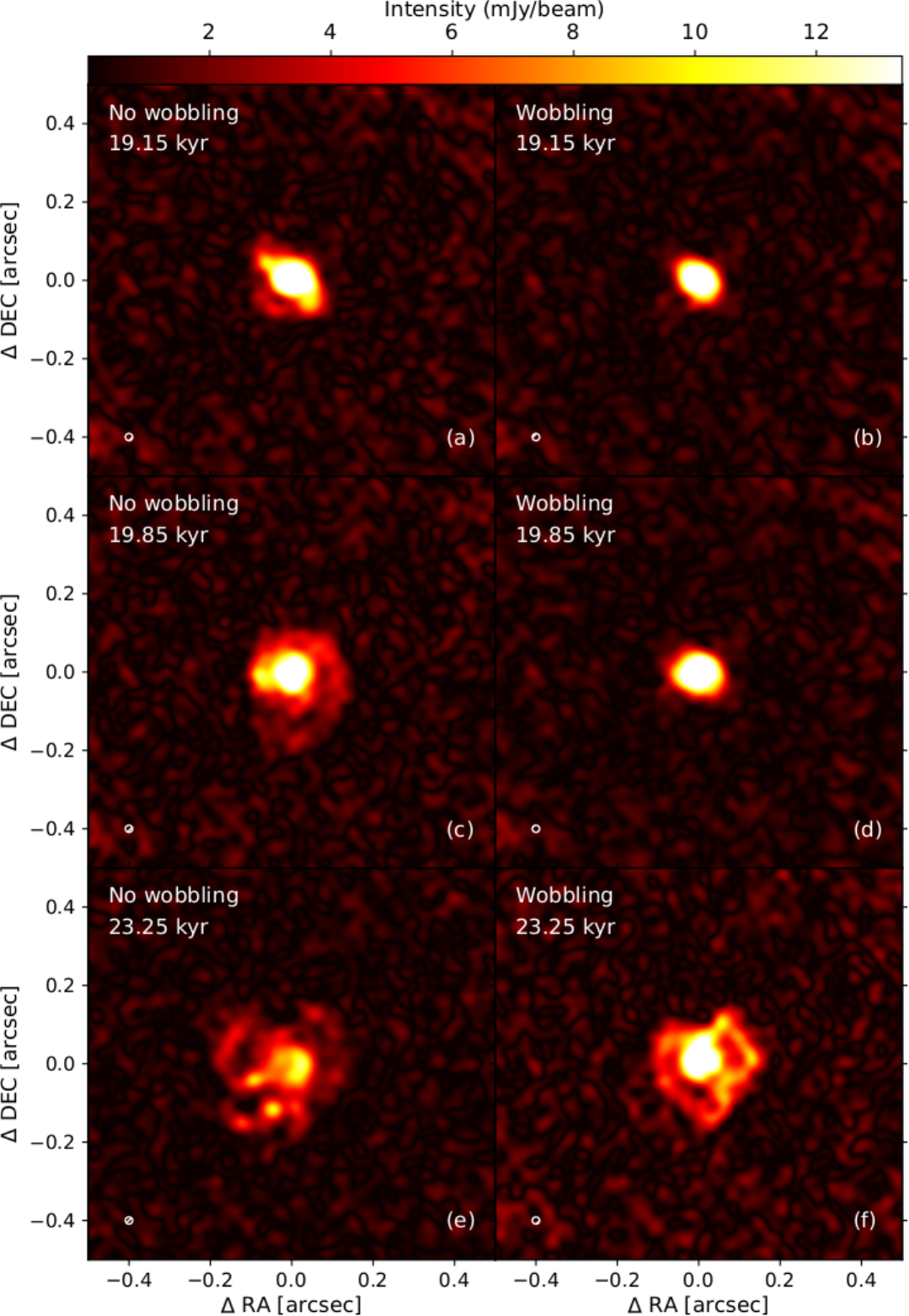}  \\
        \caption{
        Synthetic $1.2\, \rm mm$ dust continuum emission maps 
        as seen by the {\sc alma } interferometer in its antenna configuration 
        10, of the accretion disc simulated without (left panels) 
        and with (right panels) stellar inertia. The panels are shown 
        for several time instances displayed in Fig. \ref{fig:disc_density_1}, 
        namely, $19.15$~kyr (a,b), $19.85$~kyr (c,d) and $23.25\, \rm kyr$ (e,f). 
        The distance to the source is assumed to be $1 \rm kpc$.
        }
        \label{fig:disc_casa_1}  
\end{figure*}

\begin{figure*}
        \centering
        \includegraphics[width=0.82\textwidth]{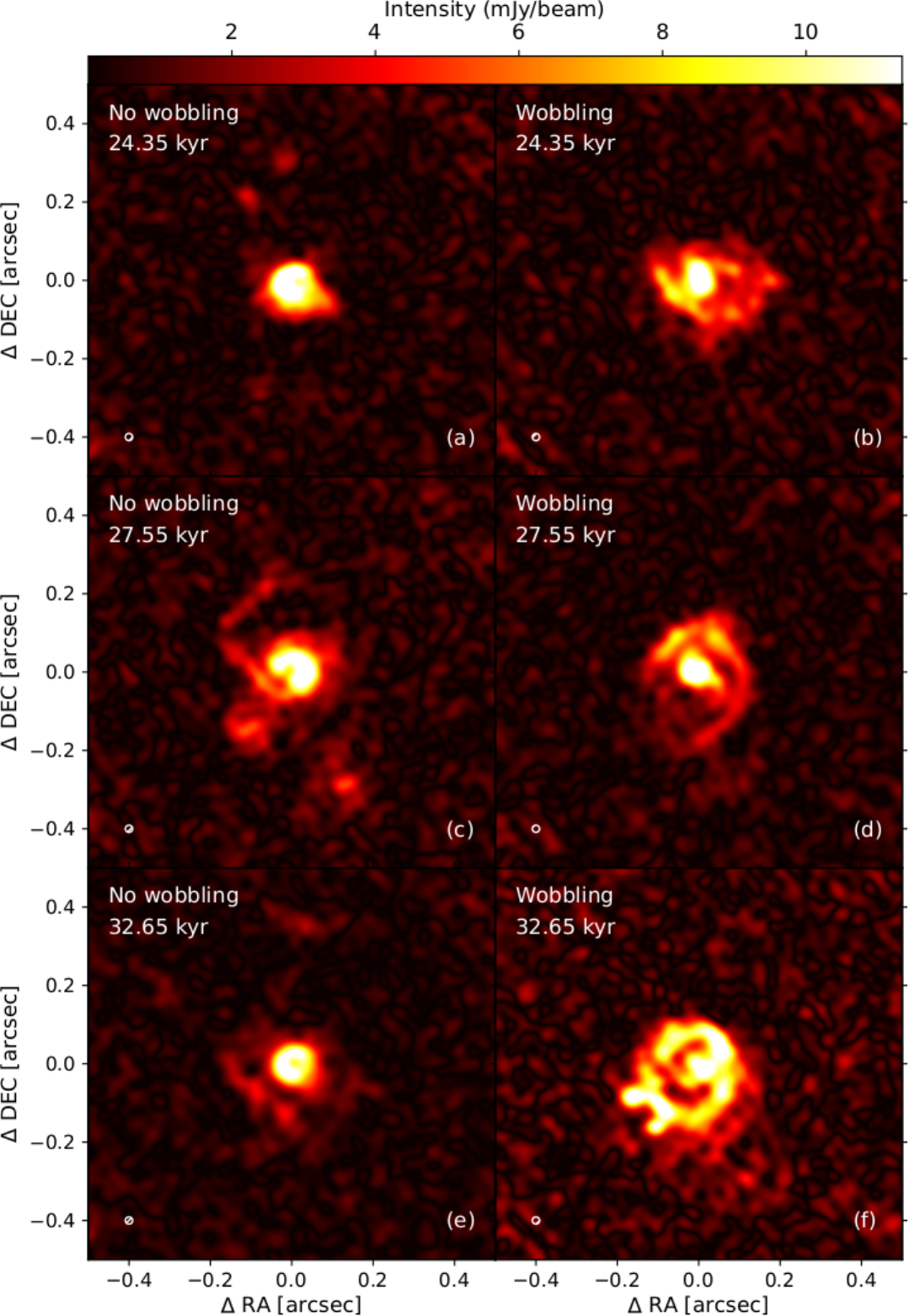}  \\
        \caption{
        As Fig. \ref{fig:disc_casa_1}, for the time instances  
        displayed in displayed in Fig. \ref{fig:disc_density_2}, 
        namely $24.35$~kyr (a,b), $27.55$~kyr (c,d) and $32.65\, \rm kyr$ (e,f). 
        }
        \label{fig:disc_casa_2}  
\end{figure*}

\section{Results}  
\label{sect:results}

This section presents the performed high-resolution disc models and details their 
accretion disc morphological properties.

\subsection{Disc evolution: $19.15-23.25\, \rm kyr$  after onset of the collapse}
\label{sect:discs_models}

Fig.\ref{fig:disc_density_1} displays the midplane density field of the accretion disc 
(in $\rm g\, \rm cm^{-3}$) in our simulations $\rm Run$$-$$512$$-$$100\rm M_{\odot}$$-$$4$\%$-\rm wio$ 
(without stellar wobbling, left-hand panels) and $\rm Run$$-$$512$$-$$100\rm M_{\odot}$$-$$4$\%$-\rm wi$ (with stellar wobbling, right-hand panels). 
The discs models are plotted at several characteristic time instances such as $19.15\, \rm kyr$ (top 
series of panels), $19.85\, \rm kyr$ (middle series of panels), $23.25\, \rm kyr$ (bottom series of 
panels) after onset of the gravitational collapse of the molecular cloud. 
The panels depict the inner ($\le 1600\, \rm au$) region of the computational domain. 
The simulation begin with the  gravitational collapse of the rotating pre-stellar 
molecular cloud and an accretion disc forms when the infalling flow hits the 
centrifugal barrier near inner computational boundary engendered by the 
conservation of angular momentum of the infalling matter.

At time $19.15\, \rm kyr$, accretion discs in both models have a round morphology with 
radii of about 
a few $100\, \rm au$. The accretion disc is already structured with several spiral arms wrapped 
around the inner region of the disc. At that early time, differences between the models 
are noticeable in the sense that the disc modelled without stellar inertia is larger by 
almost a factor $\sim 2$ and the spiral arms are more pronounced, extending to larger radii 
(Fig.\ref{fig:disc_density_1}a), while the disc in the model with stellar inertia is still smaller 
(Fig.\ref{fig:disc_density_1}b). 
A gaseous clump has formed by efficient gravitational instability in the disc without stellar 
inertia, which is the sign of faster gravitational fragmentation 
of the disc when the central protostar is not allowed to move  (Fig.\ref{fig:disc_density_1}a). 

\FloatBarrier

At time $19.85\, \rm kyr$, one can see that the disc modelled without stellar inertia keeps 
on growing in size up to a radius $\sim 450\, \rm au$. Its internal structure includes spiral arms of different sizes and thicknesses, separated 
by a more diluted interarmed space. The innermost arm is thinner, denser and hosts a 
circumstellar gaseous clump that is currently migrating downward to the inner disc region, 
where the protostar is located and accretes mass. This mechanism is responsible for the 
production of accretion-driven outbursts~\citep{meyer_mnras_464_2017} and/or the formation 
of close binary companions to the young high-mass star~\citep{meyer_mnras_473_2018}. 
Another clump, also formed by gravitational instability in the accretion disc, is ejected 
by the gravitational swing three-body mechanism away from its parent accretion disc (Fig.\ref{fig:disc_density_1}c). 
This is a major difference between the simulations without stellar inertia and those in which 
stellar inertia is included. Indeed, the disc modelled with stellar wobbling continues to display  
its round, slightly structured morphology, but it does not exhibit any sign of fragmentation 
up to this time instance (Fig.\ref{fig:disc_density_1}d). This confirms, on the basis of 
models with higher 
spatial resolution, the delay of disc fragmentation induced by stellar inertia compared 
to simulations neglecting the stellar motion in response to the disc-star gravitational interaction~\citep{meyer_mnras_517_2022}.

At $23.25\, \rm kyr$,  the model $\rm Run$$-$$512$$-$$100\rm M_{\odot}$$-$$4$\%$-\rm wio$ 
that does not include stellar inertia displays an accretion disc, which has violently fragmented and exhibits all associated features, such as several spiral arms, many gaseous 
circumstellar clumps~\citep{meyer_mnras_473_2018}, as well as a trail left behind the 
ejected clump (see Fig.\ref{fig:disc_density_1}c). We note that this clump has now travelled 
away from the disc at a distance $\ge 1600\, \rm au$ (Fig.\ref{fig:disc_density_1}e). 
The clump ejection mechanism was described for the first time 
in the context of low-mass star formation, when gravitationally fragmenting  accretion discs of  
solar-type stars generate free-floating clumps thrown through the ISM, see the studies 
of~\citet{basu_apj_750_2012,vorobyov_aa_590_2016}.  
This time instance also reveals the first clear sign of efficient gravitational instability in 
the accretion disc model simulated with stellar inertia. A characteristic disc morphology  
develops, made of several spiral arms of similar extent reaching $\ge 800\, \rm au$, 
together with the establishment of a dense disc region where both spiral arms originate and 
gaseous clumps migrate (Fig.\ref{fig:disc_density_1}f). The two numerical models differ only 
slightly  by the disc radius at this time instance, but significantly by the number of gaseous clumps in them, which is notably higher
in the case ignoring stellar inertia. This strengthens the conclusions of~\citet{meyer_mnras_517_2022} 
regarding the slow down and protracting effects of disc wobbling in the development of 
gravitational instability in accretion discs around massive protostars (Fig.\ref{fig:disc_density_1}e).

\subsection{Disc evolution: $24.35-32.65\, \rm kyr$ after onset of the collapse}
\label{sect:discs_models}

Fig.\ref{fig:disc_density_2} presents the disc evolution at later times. 
In particular, the gas spatial distributions are plotted 
at times $24.35\, \rm kyr$ (top series of panels), 
$27.55\, \rm kyr$ (middle series of panels), $32.65\, \rm kyr$ (bottom series of panels) after onset 
of the gravitational collapse of the molecular cloud. 
The disc without stellar wobbling continues fragmenting and becoming larger as it accretes 
mass, extending its denser spiral arms and forming several disc-like structures around 
the clumps that are located at large orbits ($\sim 800\, \rm au$), see 
Fig.\ref{fig:disc_density_2}a. Most important differences between the two models 
are the disc radial extend, as well as the level of fragmentation of the disc. In particular, the disc with stellar inertia included is more compact and has smaller 
clumps not surrounded yet by their own accretion discs, see Fig.\ref{fig:disc_density_2}b. 
As underlined in~\citet{meyer_mnras_517_2022} and in~\citet{regaly_aa_601_2017} in 
the context of low-mass stars, the inclusion of stellar inertia in the 
simulations permit the transfer of part of the disc angular momentum to the wobbling star. 
Consequently, the disc angular momentum is smaller compared to the case 
without stellar inertia. Since disc fragmentation takes place at larger 
radii, compacter discs are expected to fragment less~\citep{johnson_apj_597_2003}.

Later, at times $24.35-27.55\, \rm kyr$ the differences between accretion disc modelled 
with and without stellar inertia augment and become more pronounced. 
The disc with motionless protostar keeps on expanding up to a radius $\sim 1500\, \rm au$,  
and most gaseous clumps themselves are surrounded by dense disc-like structures (Fig.\ref{fig:disc_density_2}c). Those structures around clumps  
in their turn experience accretion and acquire 
mass from the disc material, while their cores become denser and hotter, potentially up 
to the temperature of molecular hydrogen dissociation, 
see~\citet{meyer_mnras_473_2018,meyer_mnras_517_2022}. 
A smaller size of the disc modelled with stellar inertia weakens the development of 
gravitational fragmentation, and a sole migrating clump
on the way to be accreted by the protostar is visible in the inner disc region 
(Fig.\ref{fig:disc_density_2}d). 
%

Finally, at time $32.65\, \rm kyr$, the disc with the fixed central young high-mass star 
has the most extended and complex morphology, composed of a pattern of spiral arms 
of different densities and lengths, interspersed with gaseous clumps, themselves potentially 
surrounded by nascent accretion discs (Fig.\ref{fig:disc_density_2}e). This picture is 
a scaled-up version of the development of massive protoplanetary discs around young solar-type 
stars, see~\citet{vorobyov_apj_719_2010}, and in the context of 
primordial star formation~\citet{vorobyov_apj_768_2013}. 
At this time instance, the accretion disc model with stellar inertia included 
exhibits clear signs of fragmentation, with a large-scale spiral arm enrolled around 
the disc structure, and several dense clumps migrate towards the inner circumstellar 
region (Fig.\ref{fig:disc_density_2}f). The disc size is similar to that of the 
disc with  fixed protostar when it started fragmenting violently 
(Fig.\ref{fig:disc_density_2}a). 
We note the still-ongoing displacement of the massive protostar away from its initial  
location at the beginning of the gravitational collapse, when it coincided with the 
geometrical centre of the pre-stellar core, at the origin of the computational domain.


\section{Observational manifestations of stellar wobbling}
\label{sect:discussion}

This section presents predictions for the appearance of the accretion discs of 
young massive stars that we simulated, as seen by the {\sc ALMA} interferometer. 
It compares the synthetic images and draws conclusions regarding the necessary 
and relevant physics to be included in models when simulating discs around massive protostars.

\subsection{Synthetic {\sc ALMA} images of the accretion discs}
\label{sect:alma_images}

Fig.\ref{fig:disc_casa_1} shows the synthetic millimetre images of the accretion discs as seen 
by the {\sc alma} facility. The plots distinguish models without (left-hand side panels) and with 
(right-hand side panels) and the  discs models are displayed at several characteristic time instances,  
namely $19.15\, \rm kyr$ (top series of panels), $19.85\, \rm kyr$ (middle series of panels), 
$23.25\, \rm kyr$ (bottom series of panels) after onset of the gravitational collapse of the 
molecular cloud. 
The panels display the inner ($\le 1600\, \rm au$) region of the computational domain, which, at a 
distance of $1\, \rm kpc$ corresponds to a region in the plane of the sky that is $1.0\, \rm arcsec$ 
wide, in both longitude and latitude, respectively. 
Both disc synthetic images at time $19.15\, \rm kyr$ have the appearance of a bright spot, 
representing the nascent accretion disc which has just formed right after the end of the free-fall 
gravitational collapse (Fig.\ref{fig:disc_casa_1}a,b). Nothing, except for two tiny protuberances, giving the image of the disc modelled with stellar inertia a slightly eared-like structure 
(Fig.\ref{fig:disc_casa_1}a), distinguish the two models (Fig.\ref{fig:disc_casa_1}b). 
We note that the images that we produce do not assume any inclination of the disc regarding 
to the plane of the sky. 
Differences arise at time $19.85\, \rm kyr$, when the disc without stellar inertia 
initiates fragmentation and develops a large spiral arm, which is enrolled all around the star and  
has a clump in it. Both substructures are observable by {\sc alma} (Fig.\ref{fig:disc_casa_1}c), 
whereas the image of the disc with stellar inertia, which is yet not fragmented, conserves its round morphology  (Fig.\ref{fig:disc_casa_1}d).

At time $23.25\, \rm kyr$, all accretion disc models have undergone gravitational 
fragmentation, however, their thermal dust emission at 
$1.3\, \rm  mm$ is quantitatively similar, despite of clear 
morphological differences of the disc density fields. In both models, the infrared 
disc nebula reveals a bright central region of global circular shape, surrounded 
by bright spots corresponding to the migrating gaseous clumps 
(Fig.\ref{fig:disc_casa_1}e,f). 
In the case of the observation of such discs, both will conclude on the presence 
of a fragmenting disc around a massive protostar. 
It is interesting to note that the flying clump catapulted from the accretion 
disc modelled without stellar inertia (Fig.\ref{fig:disc_density_1}c) is not visible 
in the infrared, as the synthetic image of Fig.\ref{fig:disc_casa_1}c,e demonstrates. 
This can be explained by the poorer numerical resolution of the outer regions of the 
computational domain compared to the inner disc region. Such a lack of 
resolution has long-time been the reason why local numerical simulations 
of massive star formation could not consistently conclude 
on disc fragmentation~\citep{Krumholz_sci_2009}.  
%
The density and temperature contrast across the ejected clump washes out as 
it enters the outer regions with progressively lower numerical resolution, making it more difficult to detect. 
This possible explanation is strengthened by the images at time 
$24.35\, \rm kyr$, still displaying two northern 
bright spots corresponding to two massive clumps (hence dense and hot) 
surrounded by accretion structures in the disc (Fig.\ref{fig:disc_density_2}a), 
while the ejected lower-mass clump is not visible in the field-of-view.

Fig.\ref{fig:disc_casa_2} plots the later evolution of the disc as seen by {\sc ALMA} at 
times $24.35\, \rm kyr$ (top series of panels), $27.55\, \rm kyr$ (middle series of panels), 
$32.65\, \rm kyr$ (bottom series of panels). 
It essentially indicates that the increasing complexity of the accretion disc simulated 
without stellar inertia does not translate entirely into the millimeter images. 
For example, the very complex circumstellar patterns of Fig.\ref{fig:disc_density_1},
\ref{fig:disc_density_2} are not fully recognisable in Fig.\ref{fig:disc_casa_2}c 
but also in Fig.\ref{fig:disc_casa_2}e. 
Nevertheless, we note that a circumstellar disc around a gaseous clump located in the southern 
part of the disc in Fig.\ref{fig:disc_casa_2}c appears as a circular, enrolled structure 
in Fig.\ref{fig:disc_casa_2}c. This indicates that the surroundings of circumstellar clumps 
that are in their turn accreting disc material, i.e. are on the way to low-mass star 
formation and can be considered as embryos of nascent secondary stars, could be observed with 
today's available facilities. 
The disc substructure are barely visible at later time $32.65\, \rm kyr$  
in the model without stellar inertia, except in the vicinity of the protostar  
(Fig.\ref{fig:disc_casa_2}e). On the other hand, the synthetic image  of 
the accretion disc modelled with stellar inertia is well traced by the emission, 
since it is denser, and therefore it has accumulated more dust mass from the infalling 
material which reemits the protostellar light more efficiently and the discs are 
easier to be observed by {\sc alma}. 
%

\subsection{Caveats of simulation method}
\label{sect:caveats}

The limitations of the methods used to produce the models presented in this study remain 
essentially the absence of the pre-stellar core magnetic field in the initial conditions, 
directly associated to the difficulty to include non-ideal effects, such as Ohmic diffusion 
and Hall effect, in the early disc formation, as well as the use of a radiation transport scheme 
restricted to non-ionizing photons. No disc wind or photoevaporation mechanism are also 
present in the model. These caveats remain sensibly similar to those already 
presented in the precedent papers of this series, see for example~\citet{meyer_mnras_473_2018}. 
The most important improvement would be to include the magnetic field in the disc simulations, 
since the photoionizing feedback is principally relevant for the physics of the bipolar 
region growing perpendicularly to the circumstellar disc. 
Only then, the construction of sophisticated simulated disc images fine-tuned 
to particular objects will be possible. 
Apart from the physical processes that are absent in our numerical model, potential 
future improvements could be the use of a computational grid without the mid-plane 
symmetry that is currently utilised in the present study and the adoption of a sink 
cell radius of value $< 20\, \rm au$. This will better capture any vertical oscillations 
of the accretion disc and study the vertical effects of non-aligned disc of massive 
protostars. 
Such update would nonetheless solely be possible at the cost of a reduced time-step 
and therefore of a dramatic increase of the computational costs of the numerical 
simulations.

\subsection{Comparison with previous numerical works}
\label{sect:comp_models}

The work of~\citet{Mignon_aa_672_2023} performs a comparison between the 
{\sc ramses} and {\sc pluto} codes regarding the initial gravitational 
collapse of a massive pre-stellar core, the first fragmentation 
era, and the disk fragmentation of the circumstellar disc based on the 
initial conditions and grid discretization of the computational domain 
corresponding to the highest resolution simulation models of~\citet{oliva_aa_644_2020}. 
Both codes include similar microphysical processes, such as radiation 
transport with direct stellar irradiation, central gravity, and self-gravity 
of the gas. Sub-grid modules are included for disc fragmentation via sink 
particles, stellar evolution, and evolution of the dust component of the 
circumstellar material. The spatial resolution is the highest ever 
achieved for calculating disc fragmentation around a nascent 
massive star~\citep{oliva_aa_644_2020}. 
Very good agreements between the two codes have been found during the early 
free-fall collapse of the pre-stellar core and the era of the first 
fragmentation. An overall satisfying match is reported in the properties 
of the disc fragments, testifying to the consistency in the radiation-hydrodynamical 
context with self-gravity of both {\sc ramses}~\citep{teyssier_aa_385_2002,fromang_aa_457_2006} 
and {\sc pluto}~\citep{mignone_apj_170_2007,migmone_apjs_198_2012}, 
despite differences in the numerical schemes used. Indeed, {\sc pluto} utilizes 
a ray-tracing method for direct irradiation with disc radiation transport by 
Flux-Limited-Diffusion~\citep{kolb_aa_559_2013}, while {\sc ramses} 
is equipped with an M1 method~\citep{rosdhal_mnras_449_2015} for protostellar 
feedback coupled to a Flux-Limited-Diffusion module~\citep{mignon_aa_635_2020} 
for disc radiation physics. {\sc ramses} operates on a Cartesian grid with adaptive 
mesh refinement, while {\sc pluto} works with a spherical grid with 
logarithmically-expanding spacing in the radial direction. 
We refer the interested reader to further details on the algorithm differences 
as well as a deep discussion on the advantages and conveniences of the 
calculation grids in~\citet{Mignon_aa_672_2023}. 

The present paper compares two models computed with the {\sc pluto} code, with an 
implementation of the microphysical processes described in~\citet{meyer_mnras_517_2022}. 
A significant difference is found between the models without and with stellar inertia, 
which was previously reported in~\citet{meyer_mnras_517_2022} based on 
lower-resolution simulations. Disc wobbling acts as an additional effect 
resulting in slowing down and prolonging the development of gravitational 
instability in the accretion disc, reducing the number and magnitude of 
the accretion-driven bursts. This is consistent with the comparison of 
the accretion rate histories in Fig. 7 of~\citet{Mignon_aa_672_2023}, 
which reports, despite a similar quiescent accretion rate, much milder 
accretion peaks onto the growing young massive star in 
the early $10\, \rm kyr$ of its existence and indicates a smaller mass 
of the circumstellar clumps forming in the disc. 
The difference in the dynamics of the disc fragments is interpreted 
as originating from discrepancies in the central star mass due to 
the stellar accretion models utilized in the two setups. 
Our results confirm the morphological difference and changing fragment 
patterns in the disc as a result of stellar wobbling, as noted by~\citet{meyer_mnras_517_2022},  
with a spatial resolution that matches that of both the {\sc ramses} 
and {\sc pluto} models in~\citet{Mignon_aa_672_2023}. We conclude that our 
results are consistent with the discrepancies noted in~\citet{Mignon_aa_672_2023} 
and propose that stellar wobbling is the factor explaining the differences  
between {\sc ramses} and {\sc pluto} results.

\subsection{Comparison with observations}
\label{sect:observations}

The following section compares the modelled accretion discs with several available 
observations of Keplerian structures around young massive stars.

\subsubsection{The disc of AFGL 4176 mm1}
\label{sect:obs_1}

\textcolor{black}{
The accretion disc surrounding the massive protostar AFGL 4176, also known 
as G308.918+0.123 or IRAS 13395-6153, as documented by~\citet{persi_aa_157_1986}, 
exhibits circumstellar characteristics~\citep{bogelund_aa_628_2019} and follows 
a Keplerian pattern of rotation~\citep{2019arXiv191109692J}. With a radius of approximately 
$1000\, \rm au$, the disc features an asymmetric spiral arm detected through 
{\sc alma} 1.2 mm emission ~\citep{2020ApJ...896...35J}.
The protostar is $\approx 25\, \rm M_\odot$, its surrounding disc is 
$\approx 12.5\, \rm M_\odot$ and it is located at a distance of 
$\approx 4.2\, \rm kpc$, 
\textcolor{black}{
see~\citet{2019arXiv191109692J,2020ApJ...896...35J}. 
The disc spiral arm  extends up to about 
}
$1000\, \rm au$ from the protostar and it has an inner 
($\le 900\, \rm au$) substructure suggesting probable 
active disc fragmentation at work in the inner part. 
In our simulations, structures of size reaching radii of 
$1000\, \rm au$ are obtained at time $\ge 24.35\, \rm kyr$ 
when disc wobbling is excluded (Fig. \ref{fig:disc_density_2}(c) and (e)) 
and when gravitational fragmentation is very active in the disc. \textcolor{black}{The disc in the model with stellar wobbling is in general more compact and features a spiral of comparable size to that of AFGL~4176 only at $t=32.65$~kyr  (Fig. \ref{fig:disc_density_2}(f)). 
However, these differences become less evident when  synthetic images are compared (Fig.~\ref{fig:disc_casa_2}) and hence the model with wobbling can also be consistent with the spiral features of  AFGL 4176.   }
}

\subsubsection{The disc of G17.64+0.16}
\label{sect:obs_2}

\textcolor{black}{
The surroundings of the massive ($45\pm 10\, \rm M_{\odot}$) O-type protostar 
G17.64+0.16 has been shown via {\sc alma} long-baseline observations at band 6 of 
$\rm H_2O$ emission to be made of a disc structure, which material is characterized
by Keplerian rotation around the central young star~\citep{maud_aa_627_2019}. 
This disc reveals a strong concentration of dust particles distributed in 
a ring-like outer region, possibly trapped by the pressure field as transitional 
discs often do in the context of low-mass protostars, as well as clear dense 
substructures in the inner disc region. 
The Toomre analysis of the $\approx 100\, \rm pc$-large disc in G17.64+0.16 shows that no 
gravitational instabilities are at work, and, hence, it is not in a state of 
violent fragmentation. This apparent stability, despite the presence of 
substructures, is puzzling and somehow challenges our results. 
The size of our accretion disc without stellar wobbling exceeds that of 
G17.64+0.16, when the first clump forms in the disc. On the other hand, such a large disc without strong fragmentation at work is found in our model 
with stellar inertia (see Fig. \ref{fig:disc_density_1}a,d). This makes stellar wobbling a necessary ingredient to reproduce the qualitative features of 
the circumstellar medium of G17.64+0.16. 
}

\textcolor{black}{
The particularly high mass of G17.64+0.16 is also not in accordance with our series 
of models, see for example fig. 2 in the study of~\citet{meyer_mnras_517_2022} 
in which the mass of the protostar does not grow above $\approx 25\, \rm M_{\odot}$. 
Since (i) a higher initial ratio of rotational-to-gravitational energy results in the formation of protostars of lower mass but discs of larger size, as demonstrated in the parameter study 
of ~\citet{meyer_mnras_500_2021}, and that (ii) the disc of G17.64+0.16 is rather 
small and young, we conclude that the initial mass distribution of the  G17.64+0.16
pre-stellar molecular core nay have been drastically different from that 
in our models. 
Fig. 2g,h of ~\citet{meyer_mnras_500_2021} shows that only molecular 
cores of initial mass $>180\, \rm M_{\odot}$ can produce protostars entering the 
high-mass regime during the core phase of free-fall gravitational collapse, while 
reaching $\approx 35\, \rm M_{\odot}$ in its direct follow-up evolutionary phase. 
We propose that G17.64+0.16 has formed via the collapse of a massive ($>180~M_\odot$) but slowly rotating core. 
}

\subsubsection{The disc of G353.2731}
\label{sect:obs_4}

\textcolor{black}{
High-resolution images with the Jansky-Very Large Array (J-VLA) and the 
Australia Telescope Compact Array (ATCA) of the region of G353.2731 at radio 
continuum and maser emission has shown that its circumstellar medium is shaped 
as an accretion disc that is faced-on regarding to our observing line-of-sight, 
\textcolor{black}{
see the study of ~\citet{motogi_apj_849_2017}. 
The mass of the protostar is constrained to be about  $10\, \rm M_{\odot}$, 
and it is surrounded by an envelope. The inner region of the envelope, 
with a radius spanning distances of approximately $\approx 100\, \rm au$, 
was interpreted in ~\citet{motogi_apj_849_2017} to be smaller than the 
typical disc sizes predicted by previous numerical simulations. 
}
Indeed, the high-resolution models including radiative 
transport and detailed stellar irradiation of~\citet{meyer_mnras_473_2018} 
show that spiral arms of such discs can easily reach $\approx 1000\, \rm au$ 
once fragmentation is at work. 
This is confirmed by our high-resolution model without stellar wobbling 
(Fig. \ref{fig:disc_density_2}c,e). Since the main effect of stellar inertia 
that we highlight in our study is to delay disc fragmentation and to produce 
discs of reduced size, we conclude that the small disc of G353.2731 is 
consistent with our work. 
}


\section{Conclusions}
\label{sect:cc}

This paper explores the effects of stellar inertia on the early development and 
fragmentation of circumstellar accretion discs around a massive protostar.  
Three-dimensional high-resolution numerical gravito-radiation-hydrodynamical 
simulations are performed with the {\sc pluto} 
code~\citep{mignone_apj_170_2007,migmone_apjs_198_2012,vaidya_apj_865_2018}.  
The models begin at the onset of the gravitational collapse of a rotating 
molecular cloud of kinetic-to-gravitational energy ratio $\beta=4\%$ and 
are followed up to a time about $32\, \rm kyr$. 
The disc models present the following two characteristic features: a high spatial 
resolution corresponding to the most highly resolved simulation using a spherical 
coordinate system~\citep{Mignon_aa_672_2023} and a careful treatment 
of the gravitational interaction at work between the disc and the central young 
high-mass star~\citep{meyer_mnras_517_2022}. 
The latter is included as the acceleration the protostar undergoes when 
interacting gravitationally with the non-axisymmteric accretion disc~\citep{Michael_2010MNRAS,hosokawa_2015,regaly_aa_601_2017}. 
The effects on the disc morphology of the resulting stellar wobbling with 
respect to the coordinate centre is evaluated by producing 
synthetic millimeter {\sc alma} observations of the star-disc system, 
which are discussed in the context of available observations.

Notable differences are found between the disc structure with and without 
stellar wobbling. When the star is allowed to move, the 
development of gravitational instabilities is slowed down and delayed. The 
disc keeps displaying a small, round morphology, in which the formation of 
spiral arms and gaseous clumps happens at later times compared to simulations 
performed without wobbling. This confirms the findings of~\citet{meyer_mnras_517_2022} 
obtained on the basis of simulations with lower-resolution. 
\textcolor{black}{ 
High-resolution accretion discs modelled with stellar inertia generate a 
much simpler pattern. It includes less gaseous clumps in it, around which secondary 
disc-like nebulae do not form as easily and numerously as in the model 
with fixed central star.  
}
Gaseous clumps ejected from the disc in the model without stellar wobbling 
~\citet{basu_apj_750_2012,vorobyov_aa_590_2016} are not found once stellar 
inertia is included into the model, at least for the kinetic-to-gravitational 
energy ratio that we consider ($\beta=4\%$).

Synthetic $1.2\, \rm mm$ images of thermal dust emission 
of circumstellar discs wherein stellar wobbling is considered turn out to be much more 
qualitatively consistent with real observations of the discs of massive protostars than the images produced in models without stellar inertia, in which the star is fixed to the coordinate origin of the domain.   
\textcolor{black}{
This applies to the observations of the Keplerian circumstellar medium of
AFGL 4176 mm1, G17.64+0.16 and  G353.273. 
}
This study stresses the importance of spatial resolution in the study of accretion 
discs around massive protostars and it demonstrates that including the star-to-disc 
gravitational interaction is a preponderant ingredient to account for in 
realistic modelling of the surroundings of young massive stellar objects.


\begin{acknowledgements}
We are thankful to the anonymous referee for comments that helped to improve the manuscript.
The author acknowledge the North-German Supercomputing Alliance (HLRN) for providing 
HPC resources that have contributed to the research results reported in this paper. 
This research made use of the {\sc pluto} code developed at the University of Torino  
by A.~Mignone (http://plutocode.ph.unito.it/)  
and of the {\sc radmc-3d} code developed at the University of Heidelberg by C.~Dullemond 
(https://www.ita.uni-heidelberg.de/$\sim$dullemond/software/radmc-3d/).
The figures have been produced using the Matplotlib plotting library for the 
Python programming language (https://matplotlib.org/). 
The data underlying this article will be shared on reasonable request to the 
corresponding author. 
This work has been supported by the grant PID2021-124581OB-I00 funded by 
MCIN/AEI/10.13039/501100011033 and 2021SGR00426 of the Generalitat de Catalunya. 
This work was also supported by the Spanish program Unidad de Excelencia Mar\' ia 
de Maeztu CEX2020-001058-M.
This work also supported by MCIN with funding from European Union 
NextGeneration EU (PRTR-C17.I1). E.I.V. acknowledges support from the Russian Science Foundation, project No. 23-12-00258. 
\end{acknowledgements}


\bibliographystyle{aa} 
\bibliography{grid} 

\begin{thebibliography}{112}
\expandafter\ifx\csname natexlab\endcsname\relax\def\natexlab#1{#1}\fi

\bibitem[{{Ahmadi} {et~al.}(2018){Ahmadi}, {Beuther}, {Mottram}, {Bosco},
  {Linz}, {Henning}, {Winters}, {Kuiper}, {Pudritz}, {S{\'a}nchez-Monge},
  {Keto}, {Beltran}, {Bontemps}, {Cesaroni}, {Csengeri}, {Feng}, \&
  {Galvan-Madrid}}]{ahmadi_aa_618_2018}
{Ahmadi}, A., {Beuther}, H., {Mottram}, J.~C., {et~al.} 2018, \aap, 618, A46

\bibitem[{{Ahmadi} {et~al.}(2019){Ahmadi}, {Kuiper}, \&
  {Beuther}}]{ahmadi_aa_632_2019}
{Ahmadi}, A., {Kuiper}, R., \& {Beuther}, H. 2019, \aap, 632, A50

\bibitem[{{Basu} \& {Vorobyov}(2012)}]{basu_apj_750_2012}
{Basu}, S. \& {Vorobyov}, E.~I. 2012, \apj, 750, 30

\bibitem[{{Bitsch} {et~al.}(2014){Bitsch}, {Morbidelli}, {Lega}, \&
  {Crida}}]{bitsch_aa_564_2014}
{Bitsch}, B., {Morbidelli}, A., {Lega}, E., \& {Crida}, A. 2014, \aap, 564,
  A135

\bibitem[{{Bjorkman} \& {Wood}(2001)}]{bjorkman_apj_554_2001}
{Bjorkman}, J.~E. \& {Wood}, K. 2001, \apj, 554, 615

\bibitem[{{B{\o}gelund} {et~al.}(2019){B{\o}gelund}, {Barr}, {Taquet},
  {Ligterink}, {Persson}, {Hogerheijde}, \& {van
  Dishoeck}}]{bogelund_aa_628_2019}
{B{\o}gelund}, E.~G., {Barr}, A.~G., {Taquet}, V., {et~al.} 2019, \aap, 628, A2

\bibitem[{{Burns}(2018)}]{burns_iaus_336_2018}
{Burns}, R.~A. 2018, in IAU Symposium, Vol. 336, Astrophysical Masers:
  Unlocking the Mysteries of the Universe, ed. A.~{Tarchi}, M.~J. {Reid}, \&
  P.~{Castangia}, 263--266

\bibitem[{{Burns} {et~al.}(2017){Burns}, {Handa}, {Imai}, {Nagayama},
  {Omodaka}, {Hirota}, {Motogi}, {van Langevelde}, \&
  {Baan}}]{burns_mnras_467_2017}
{Burns}, R.~A., {Handa}, T., {Imai}, H., {et~al.} 2017, \mnras, 467, 2367

\bibitem[{{Caratti o Garatti} {et~al.}(2015){Caratti o Garatti}, {Stecklum},
  {Linz}, {Garcia Lopez}, \& {Sanna}}]{caratti_aa_573_2015}
{Caratti o Garatti}, A., {Stecklum}, B., {Linz}, H., {Garcia Lopez}, R., \&
  {Sanna}, A. 2015, \aap, 573, A82

\bibitem[{{Cesaroni} {et~al.}(2010){Cesaroni}, {Hofner}, {Araya}, \&
  {Kurtz}}]{cesaroni_aa_509_2010}
{Cesaroni}, R., {Hofner}, P., {Araya}, E., \& {Kurtz}, S. 2010, \aap, 509, A50

\bibitem[{{Chen} {et~al.}(2017){Chen}, {Ren}, {Zhang}, {Shen}, \&
  {Qiu}}]{chen_apj_835_2017}
{Chen}, X., {Ren}, Z., {Zhang}, Q., {Shen}, Z., \& {Qiu}, K. 2017, \apj, 835,
  227

\bibitem[{{Chini} {et~al.}(2012){Chini}, {Hoffmeister}, {Nasseri}, {Stahl}, \&
  {Zinnecker}}]{chini_424_mnras_2012}
{Chini}, R., {Hoffmeister}, V.~H., {Nasseri}, A., {Stahl}, O., \& {Zinnecker},
  H. 2012, \mnras, 424, 1925

\bibitem[{{Commer{\c c}on} {et~al.}(2011){Commer{\c c}on}, {Teyssier}, {Audit},
  {Hennebelle}, \& {Chabrier}}]{commercon_aa_529_2011}
{Commer{\c c}on}, B., {Teyssier}, R., {Audit}, E., {Hennebelle}, P., \&
  {Chabrier}, G. 2011, \aap, 529, A35

\bibitem[{{Commer{\c{c}}on} {et~al.}(2022){Commer{\c{c}}on}, {Gonz{\'a}lez},
  {Mignon-Risse}, {Hennebelle}, \& {Vaytet}}]{commercon_aa_658_2022}
{Commer{\c{c}}on}, B., {Gonz{\'a}lez}, M., {Mignon-Risse}, R., {Hennebelle},
  P., \& {Vaytet}, N. 2022, \aap, 658, A52

\bibitem[{{Commer{\c{c}}on} {et~al.}(2011){Commer{\c{c}}on}, {Hennebelle}, \&
  {Henning}}]{commercon_apj_742_2011}
{Commer{\c{c}}on}, B., {Hennebelle}, P., \& {Henning}, T. 2011, \apjl, 742, L9

\bibitem[{{Cunningham} {et~al.}(2009){Cunningham}, {Moeckel}, \&
  {Bally}}]{Cunningham_apj_692_2009}
{Cunningham}, N.~J., {Moeckel}, N., \& {Bally}, J. 2009, \apj, 692, 943

\bibitem[{{Dong} {et~al.}(2016){Dong}, {Vorobyov}, {Pavlyuchenkov}, {Chiang},
  \& {Liu}}]{dong_apj_823_2016}
{Dong}, R., {Vorobyov}, E., {Pavlyuchenkov}, Y., {Chiang}, E., \& {Liu}, H.~B.
  2016, \apj, 823, 141

\bibitem[{{Dullemond}(2012)}]{dullemond_2012}
{Dullemond}, C.~P. 2012, {RADMC-3D: A multi-purpose radiative transfer tool},
  Astrophysics Source Code Library

\bibitem[{{Elbakyan} {et~al.}(2023){Elbakyan}, {Nayakshin}, {Meyer}, \&
  {Vorobyov}}]{meyer_mnras_518_2023}
{Elbakyan}, V.~G., {Nayakshin}, S., {Meyer}, D. M.~A., \& {Vorobyov}, E.~I.
  2023, \mnras, 518, 791

\bibitem[{{Elbakyan} {et~al.}(2021){Elbakyan}, {Nayakshin}, {Vorobyov},
  {Caratti o Garatti}, \& {Eisl{\"o}ffel}}]{elbakyan_aa_651_2021}
{Elbakyan}, V.~G., {Nayakshin}, S., {Vorobyov}, E.~I., {Caratti o Garatti}, A.,
  \& {Eisl{\"o}ffel}, J. 2021, \aap, 651, L3

\bibitem[{{Elbakyan} {et~al.}(2019){Elbakyan}, {Vorobyov}, {Rab}, {Meyer},
  {G{\"u}del}, {Hosokawa}, \& {Yorke}}]{elbakyan_mnras_484_2019}
{Elbakyan}, V.~G., {Vorobyov}, E.~I., {Rab}, C., {et~al.} 2019, \mnras, 484,
  146

\bibitem[{{Flock} {et~al.}(2013){Flock}, {Fromang}, {Gonz{\'a}lez}, \&
  {Commer{\c c}on}}]{flock_aa_560_2013}
{Flock}, M., {Fromang}, S., {Gonz{\'a}lez}, M., \& {Commer{\c c}on}, B. 2013,
  \aap, 560, A43

\bibitem[{{Forgan} {et~al.}(2016){Forgan}, {Ilee}, {Cyganowski}, {Brogan}, \&
  {Hunter}}]{forgan_mnras_463_2016}
{Forgan}, D.~H., {Ilee}, J.~D., {Cyganowski}, C.~J., {Brogan}, C.~L., \&
  {Hunter}, T.~R. 2016, \mnras, 463, 957

\bibitem[{{Fromang} {et~al.}(2006){Fromang}, {Hennebelle}, \&
  {Teyssier}}]{fromang_aa_457_2006}
{Fromang}, S., {Hennebelle}, P., \& {Teyssier}, R. 2006, \aap, 457, 371

\bibitem[{{Ginsburg} {et~al.}(2018){Ginsburg}, {Bally}, {Goddi}, {Plambeck}, \&
  {Wright}}]{2018arXiv180410622G}
{Ginsburg}, A., {Bally}, J., {Goddi}, C., {Plambeck}, R., \& {Wright}, M. 2018,
  \apj, 860, 119

\bibitem[{{Guadarrama} {et~al.}(2023){Guadarrama}, {Vorobyov}, {Rab},
  {G{\"u}del}, {Garatti}, \& {Sobolev}}]{2023arXiv231207184G}
{Guadarrama}, R., {Vorobyov}, E., {Rab}, C., {et~al.} 2023, arXiv e-prints,
  arXiv:2312.07184

\bibitem[{{Harries}(2015)}]{harries_mnras_448_2015}
{Harries}, T.~J. 2015, \mnras, 448, 3156

\bibitem[{{Harries} {et~al.}(2017){Harries}, {Douglas}, \&
  {Ali}}]{harries_2017}
{Harries}, T.~J., {Douglas}, T.~A., \& {Ali}, A. 2017, \mnras, 471, 4111

\bibitem[{{Hennebelle} {et~al.}(2016){Hennebelle}, {Commer{\c c}on},
  {Chabrier}, \& {Marchand}}]{hennebelle_apj_830_2016}
{Hennebelle}, P., {Commer{\c c}on}, B., {Chabrier}, G., \& {Marchand}, P. 2016,
  \apjl, 830, L8

\bibitem[{{Hirano} {et~al.}(2017){Hirano}, {Hosokawa}, {Yoshida}, \&
  {Kuiper}}]{Hirano_2017Science}
{Hirano}, S., {Hosokawa}, T., {Yoshida}, N., \& {Kuiper}, R. 2017, Science,
  357, 1375

\bibitem[{{Hosokawa} {et~al.}(2016){Hosokawa}, {Hirano}, {Kuiper}, {Yorke},
  {Omukai}, \& {Yoshida}}]{hosokawa_2015}
{Hosokawa}, T., {Hirano}, S., {Kuiper}, R., {et~al.} 2016, \apj, 824, 119

\bibitem[{{Hosokawa} \& {Omukai}(2009)}]{hosokawa_apj_691_2009}
{Hosokawa}, T. \& {Omukai}, K. 2009, \apj, 691, 823

\bibitem[{{Hosokawa} {et~al.}(2010){Hosokawa}, {Yorke}, \&
  {Omukai}}]{hosokawa_apj_721_2010}
{Hosokawa}, T., {Yorke}, H.~W., \& {Omukai}, K. 2010, \apj, 721, 478

\bibitem[{{Ilee} {et~al.}(2018){Ilee}, {Cyganowski}, {Brogan}, {Hunter},
  {Forgan}, {Haworth}, {Clarke}, \& {Harries}}]{2018ApJ...869L..24I}
{Ilee}, J.~D., {Cyganowski}, C.~J., {Brogan}, C.~L., {et~al.} 2018, \apjl, 869,
  L24

\bibitem[{{Ilee} {et~al.}(2016){Ilee}, {Cyganowski}, {Nazari}, {Hunter},
  {Brogan}, {Forgan}, \& {Zhang}}]{ilee_mnras_462_2016}
{Ilee}, J.~D., {Cyganowski}, C.~J., {Nazari}, P., {et~al.} 2016, \mnras, 462,
  4386

\bibitem[{{Jankovic} {et~al.}(2019){Jankovic}, {Haworth}, {Ilee}, {Forgan},
  {Cyganowski}, {Walsh}, {Brogan}, {Hunter}, \&
  {Mohanty}}]{jankovic_mnras_482_2019}
{Jankovic}, M.~R., {Haworth}, T.~J., {Ilee}, J.~D., {et~al.} 2019, \mnras, 482,
  4673

\bibitem[{{Johnson} \& {Gammie}(2003)}]{johnson_apj_597_2003}
{Johnson}, B.~M. \& {Gammie}, C.~F. 2003, \apj, 597, 131

\bibitem[{{Johnston} {et~al.}(2020{\natexlab{a}}){Johnston}, {Hoare},
  {Beuther}, {Linz}, {Boley}, {Kuiper}, {Kee}, \&
  {Robitaille}}]{2019arXiv191109692J}
{Johnston}, K.~G., {Hoare}, M.~G., {Beuther}, H., {et~al.} 2020{\natexlab{a}},
  \apj, 896, 35

\bibitem[{{Johnston} {et~al.}(2020{\natexlab{b}}){Johnston}, {Hoare},
  {Beuther}, {Linz}, {Boley}, {Kuiper}, {Kee}, \&
  {Robitaille}}]{2020ApJ...896...35J}
{Johnston}, K.~G., {Hoare}, M.~G., {Beuther}, H., {et~al.} 2020{\natexlab{b}},
  \apj, 896, 35

\bibitem[{{Johnston} {et~al.}(2015){Johnston}, {Robitaille}, {Beuther}, {Linz},
  {Boley}, {Kuiper}, {Keto}, {Hoare}, \& {van Boekel}}]{johnston_apj_813_2015}
{Johnston}, K.~G., {Robitaille}, T.~P., {Beuther}, H., {et~al.} 2015, \apjl,
  813, L19

\bibitem[{{Keto} \& {Wood}(2006)}]{keto_apj_637_2006}
{Keto}, E. \& {Wood}, K. 2006, \apj, 637, 850

\bibitem[{{Klassen} {et~al.}(2014){Klassen}, {Kuiper}, {Pudritz}, {Peters},
  {Banerjee}, \& {Buntemeyer}}]{klassen_apj_797_2014}
{Klassen}, M., {Kuiper}, R., {Pudritz}, R.~E., {et~al.} 2014, \apj, 797, 4

\bibitem[{{Klassen} {et~al.}(2016){Klassen}, {Pudritz}, {Kuiper}, {Peters}, \&
  {Banerjee}}]{klassen_apj_823_2016}
{Klassen}, M., {Pudritz}, R.~E., {Kuiper}, R., {Peters}, T., \& {Banerjee}, R.
  2016, \apj, 823, 28

\bibitem[{{Kobulnicky} {et~al.}(2014){Kobulnicky}, {Kiminki}, {Lundquist},
  {Burke}, {Chapman}, {Keller}, {Lester}, {Rolen}, {Topel}, {Bhattacharjee},
  {Smullen}, {Vargas {\'A}lvarez}, {Runnoe}, {Dale}, \&
  {Brotherton}}]{2014ApJS..213...34K}
{Kobulnicky}, H.~A., {Kiminki}, D.~C., {Lundquist}, M.~J., {et~al.} 2014,
  \apjs, 213, 34

\bibitem[{{Kolb} {et~al.}(2013){Kolb}, {Stute}, {Kley}, \&
  {Mignone}}]{kolb_aa_559_2013}
{Kolb}, S.~M., {Stute}, M., {Kley}, W., \& {Mignone}, A. 2013, \aap, 559, A80

\bibitem[{{Kraus} {et~al.}(2017){Kraus}, {Kluska}, {Kreplin}, {Bate},
  {Harries}, {Hofmann}, {Hone}, {Monnier}, {Weigelt}, {Anugu}, {de Wit}, \&
  {Wittkowski}}]{kraus_apj_835_2017}
{Kraus}, S., {Kluska}, J., {Kreplin}, A., {et~al.} 2017, \apjl, 835, L5

\bibitem[{{Krumholz} {et~al.}(2007{\natexlab{a}}){Krumholz}, {Klein}, \&
  {McKee}}]{krumholz_apj_665_2007}
{Krumholz}, M.~R., {Klein}, R.~I., \& {McKee}, C.~F. 2007{\natexlab{a}}, \apj,
  665, 478

\bibitem[{{Krumholz} {et~al.}(2007{\natexlab{b}}){Krumholz}, {Klein}, \&
  {McKee}}]{krumholz_apj_656_2007}
{Krumholz}, M.~R., {Klein}, R.~I., \& {McKee}, C.~F. 2007{\natexlab{b}}, \apj,
  656, 959

\bibitem[{{Krumholz} {et~al.}(2009{\natexlab{a}}){Krumholz}, {Klein}, {McKee},
  {Offner}, \& {Cunningham}}]{krumholz_sci_323_2009}
{Krumholz}, M.~R., {Klein}, R.~I., {McKee}, C.~F., {Offner}, S.~S.~R., \&
  {Cunningham}, A.~J. 2009{\natexlab{a}}, Science, 323, 754

\bibitem[{{Krumholz} {et~al.}(2009{\natexlab{b}}){Krumholz}, {Klein}, {McKee},
  {Offner}, \& {Cunningham}}]{Krumholz_sci_2009}
{Krumholz}, M.~R., {Klein}, R.~I., {McKee}, C.~F., {Offner}, S.~S.~R., \&
  {Cunningham}, A.~J. 2009{\natexlab{b}}, Science, 323, 754

\bibitem[{{Kuiper} {et~al.}(2010{\natexlab{a}}){Kuiper}, {Klahr}, {Beuther}, \&
  {Henning}}]{kuiper_apj_722_2010}
{Kuiper}, R., {Klahr}, H., {Beuther}, H., \& {Henning}, T. 2010{\natexlab{a}},
  \apj, 722, 1556

\bibitem[{{Kuiper} {et~al.}(2011){Kuiper}, {Klahr}, {Beuther}, \&
  {Henning}}]{kuiper_apj_732_2011}
{Kuiper}, R., {Klahr}, H., {Beuther}, H., \& {Henning}, T. 2011, \apj, 732, 20

\bibitem[{{Kuiper} {et~al.}(2010{\natexlab{b}}){Kuiper}, {Klahr}, {Dullemond},
  {Kley}, \& {Henning}}]{kuiper_aa_511_2010}
{Kuiper}, R., {Klahr}, H., {Dullemond}, C., {Kley}, W., \& {Henning}, T.
  2010{\natexlab{b}}, \aap, 511, A81

\bibitem[{{Kuiper} \& {Yorke}(2013{\natexlab{a}})}]{kuiper_apj_763_2013}
{Kuiper}, R. \& {Yorke}, H.~W. 2013{\natexlab{a}}, \apj, 763, 104

\bibitem[{{Kuiper} \& {Yorke}(2013{\natexlab{b}})}]{kuiper_apj_772_2013}
{Kuiper}, R. \& {Yorke}, H.~W. 2013{\natexlab{b}}, \apj, 772, 61

\bibitem[{{Laor} \& {Draine}(1993)}]{laor_apj_402_1993}
{Laor}, A. \& {Draine}, B.~T. 1993, \apj, 402, 441

\bibitem[{Machida {et~al.}(2011)Machida, ichiro Inutsuka, \&
  Matsumoto}]{Machida_2011}
Machida, M.~N., ichiro Inutsuka, S., \& Matsumoto, T. 2011, The Astrophysical
  Journal, 729, 42

\bibitem[{{Mahy} {et~al.}(2013){Mahy}, {Rauw}, {De Becker}, {Eenens}, \&
  {Flores}}]{2013A&A...550A..27M}
{Mahy}, L., {Rauw}, G., {De Becker}, M., {Eenens}, P., \& {Flores}, C.~A. 2013,
  \aap, 550, A27

\bibitem[{{Maud} {et~al.}(2019){Maud}, {Cesaroni}, {Kumar}, {Rivilla},
  {Ginsburg}, {Klaassen}, {Harsono}, {S{\'a}nchez-Monge}, {Ahmadi}, {Allen},
  {Beltr{\'a}n}, {Beuther}, {Galv{\'a}n-Madrid}, {Goddi}, {Hoare},
  {Hogerheijde}, {Johnston}, {Kuiper}, {Moscadelli}, {Peters}, {Testi}, {van
  der Tak}, \& {de Wit}}]{maud_aa_627_2019}
{Maud}, L.~T., {Cesaroni}, R., {Kumar}, M.~S.~N., {et~al.} 2019, \aap, 627, L6

\bibitem[{{Maud} {et~al.}(2018){Maud}, {Cesaroni}, {Kumar}, {van der Tak},
  {Allen}, {Hoare}, {Klaassen}, {Harsono}, \& {Hogerheijde}}]{maud_aa_620_2018}
{Maud}, L.~T., {Cesaroni}, R., {Kumar}, M.~S.~N., {et~al.} 2018, \aap, 620, A31

\bibitem[{{Maud} {et~al.}(2017){Maud}, {Hoare}, {Galv{\'a}n-Madrid}, {Zhang},
  {de Wit}, {Keto}, {Johnston}, \& {Pineda}}]{maud_467_mnras_2017}
{Maud}, L.~T., {Hoare}, M.~G., {Galv{\'a}n-Madrid}, R., {et~al.} 2017, \mnras,
  467, L120

\bibitem[{{McMullin} {et~al.}(2007){McMullin}, {Waters}, {Schiebel}, {Young},
  \& {Golap}}]{McMullin_aspc_376_2007}
{McMullin}, J.~P., {Waters}, B., {Schiebel}, D., {Young}, W., \& {Golap}, K.
  2007, in Astronomical Society of the Pacific Conference Series, Vol. 376,
  Astronomical Data Analysis Software and Systems XVI, ed. R.~A. {Shaw},
  F.~{Hill}, \& D.~J. {Bell}, 127

\bibitem[{{Meyer} {et~al.}(2019{\natexlab{a}}){Meyer}, {Haemmerl{\'e}}, \&
  {Vorobyov}}]{meyer_mnras_484_2019}
{Meyer}, D.~M.~A., {Haemmerl{\'e}}, L., \& {Vorobyov}, E.~I.
  2019{\natexlab{a}}, \mnras, 484, 2482

\bibitem[{{Meyer} {et~al.}(2019{\natexlab{b}}){Meyer}, {Kreplin}, {Kraus},
  {Vorobyov}, {Haemmerle}, \& {Eisl{\"o}ffel}}]{meyer_487_MNRAS_2019}
{Meyer}, D.~M.~A., {Kreplin}, A., {Kraus}, S., {et~al.} 2019{\natexlab{b}},
  \mnras, 487, 4473

\bibitem[{{Meyer} {et~al.}(2018){Meyer}, {Kuiper}, {Kley}, {Johnston}, \&
  {Vorobyov}}]{meyer_mnras_473_2018}
{Meyer}, D.~M.-A., {Kuiper}, R., {Kley}, W., {Johnston}, K.~G., \& {Vorobyov},
  E. 2018, \mnras, 473, 3615

\bibitem[{{Meyer} {et~al.}(2021){Meyer}, {Vorobyov}, {Elbakyan},
  {Eisl{\"o}ffel}, {Sobolev}, \& {St{\"o}hr}}]{meyer_mnras_500_2021}
{Meyer}, D.~M.~A., {Vorobyov}, E.~I., {Elbakyan}, V.~G., {et~al.} 2021, \mnras,
  500, 4448

\bibitem[{{Meyer} {et~al.}(2022){Meyer}, {Vorobyov}, {Elbakyan}, {Kraus},
  {Liu}, {Nayakshin}, \& {Sobolev}}]{meyer_mnras_517_2022}
{Meyer}, D.~M.~A., {Vorobyov}, E.~I., {Elbakyan}, V.~G., {et~al.} 2022, \mnras,
  517, 4795

\bibitem[{{Meyer} {et~al.}(2019{\natexlab{c}}){Meyer}, {Vorobyov}, {Elbakyan},
  {Stecklum}, {Eisl{\"o}ffel}, \& {Sobolev}}]{meyer_mnras_482_2019}
{Meyer}, D.~M.-A., {Vorobyov}, E.~I., {Elbakyan}, V.~G., {et~al.}
  2019{\natexlab{c}}, \mnras, 482, 5459

\bibitem[{{Meyer} {et~al.}(2017){Meyer}, {Vorobyov}, {Kuiper}, \&
  {Kley}}]{meyer_mnras_464_2017}
{Meyer}, D.~M.-A., {Vorobyov}, E.~I., {Kuiper}, R., \& {Kley}, W. 2017, \mnras,
  464, L90

\bibitem[{{Michael} \& {Durisen}(2010)}]{Michael_2010MNRAS}
{Michael}, S. \& {Durisen}, R.~H. 2010, \mnras, 406, 279

\bibitem[{{Mignon-Risse} {et~al.}(2023{\natexlab{a}}){Mignon-Risse},
  {Gonz{\'a}lez}, \& {Commer{\c{c}}on}}]{mignonrisse_673_aa_2023}
{Mignon-Risse}, R., {Gonz{\'a}lez}, M., \& {Commer{\c{c}}on}, B.
  2023{\natexlab{a}}, \aap, 673, A134

\bibitem[{{Mignon-Risse} {et~al.}(2020){Mignon-Risse}, {Gonz{\'a}lez},
  {Commer{\c{c}}on}, \& {Rosdahl}}]{mignon_aa_635_2020}
{Mignon-Risse}, R., {Gonz{\'a}lez}, M., {Commer{\c{c}}on}, B., \& {Rosdahl}, J.
  2020, \aap, 635, A42

\bibitem[{{Mignon-Risse} {et~al.}(2021){Mignon-Risse}, {Gonz{\'a}lez},
  {Commer{\c{c}}on}, \& {Rosdahl}}]{mignon_risse_aa_652_2021}
{Mignon-Risse}, R., {Gonz{\'a}lez}, M., {Commer{\c{c}}on}, B., \& {Rosdahl}, J.
  2021, \aap, 652, A69

\bibitem[{{Mignon-Risse} {et~al.}(2023{\natexlab{b}}){Mignon-Risse}, {Oliva},
  {Gonz{\'a}lez}, {Kuiper}, \& {Commer{\c{c}}on}}]{Mignon_aa_672_2023}
{Mignon-Risse}, R., {Oliva}, A., {Gonz{\'a}lez}, M., {Kuiper}, R., \&
  {Commer{\c{c}}on}, B. 2023{\natexlab{b}}, \aap, 672, A88

\bibitem[{{Mignone} {et~al.}(2007){Mignone}, {Bodo}, {Massaglia}, {Matsakos},
  {Tesileanu}, {Zanni}, \& {Ferrari}}]{mignone_apj_170_2007}
{Mignone}, A., {Bodo}, G., {Massaglia}, S., {et~al.} 2007, \apjs, 170, 228

\bibitem[{{Mignone} {et~al.}(2012){Mignone}, {Zanni}, {Tzeferacos}, {van
  Straalen}, {Colella}, \& {Bodo}}]{migmone_apjs_198_2012}
{Mignone}, A., {Zanni}, C., {Tzeferacos}, P., {et~al.} 2012, \apjs, 198, 7

\bibitem[{{Motogi} {et~al.}(2017){Motogi}, {Hirota}, {Sorai}, {Yonekura},
  {Sugiyama}, {Honma}, {Niinuma}, {Hachisuka}, {Fujisawa}, \&
  {Walsh}}]{motogi_apj_849_2017}
{Motogi}, K., {Hirota}, T., {Sorai}, K., {et~al.} 2017, \apj, 849, 23

\bibitem[{{Nayakshin}(2010)}]{2010MNRAS.408L..36N}
{Nayakshin}, S. 2010, \mnras, 408, L36

\bibitem[{{Nayakshin}(2016)}]{nayakshin_mnras_461_2016}
{Nayakshin}, S. 2016, \mnras, 461, 3194

\bibitem[{{Nayakshin}(2017)}]{2017PASA...34....2N}
{Nayakshin}, S. 2017, \pasa, 34, e002

\bibitem[{{Nayakshin} \& {Lodato}(2012)}]{nayakshin_mnras_426_2012}
{Nayakshin}, S. \& {Lodato}, G. 2012, \mnras, 426, 70

\bibitem[{{Oliva} \& {Kuiper}(2023{\natexlab{a}})}]{oliva_aa_669_2023}
{Oliva}, A. \& {Kuiper}, R. 2023{\natexlab{a}}, \aap, 669, A80

\bibitem[{{Oliva} \& {Kuiper}(2023{\natexlab{b}})}]{oliva_aa_669_II_2023}
{Oliva}, A. \& {Kuiper}, R. 2023{\natexlab{b}}, \aap, 669, A81

\bibitem[{{Oliva} \& {Kuiper}(2020)}]{oliva_aa_644_2020}
{Oliva}, G.~A. \& {Kuiper}, R. 2020, \aap, 644, A41

\bibitem[{{Papaloizou} \& {Savonije}(1991)}]{papaloizou_mnras_248_1991}
{Papaloizou}, J.~C. \& {Savonije}, G.~J. 1991, \mnras, 248, 353

\bibitem[{{Persi} {et~al.}(1986){Persi}, {Ferrari-Toniolo}, \&
  {Spinoglio}}]{persi_aa_157_1986}
{Persi}, P., {Ferrari-Toniolo}, M., \& {Spinoglio}, L. 1986, \aap, 157, 29

\bibitem[{{Peters} {et~al.}(2010){Peters}, {Banerjee}, {Klessen}, {Mac Low},
  {Galv{\'a}n-Madrid}, \& {Keto}}]{peters_apj_711_2010}
{Peters}, T., {Banerjee}, R., {Klessen}, R.~S., {et~al.} 2010, \apj, 711, 1017

\bibitem[{{Pickett} {et~al.}(2003){Pickett}, {Mej{\'\i}a}, {Durisen}, {Cassen},
  {Berry}, \& {Link}}]{pickett_apj_590_2003}
{Pickett}, B.~K., {Mej{\'\i}a}, A.~C., {Durisen}, R.~H., {et~al.} 2003, \apj,
  590, 1060

\bibitem[{{Purser} {et~al.}(2018){Purser}, {Lumsden}, {Hoare}, \&
  {Cunningham}}]{purser_mnras_475_2018}
{Purser}, S.~J.~D., {Lumsden}, S.~L., {Hoare}, M.~G., \& {Cunningham}, N. 2018,
  \mnras, 475, 2

\bibitem[{{Purser} {et~al.}(2016){Purser}, {Lumsden}, {Hoare}, {Urquhart},
  {Cunningham}, {Purcell}, {Brooks}, {Garay}, {G{\'u}zman}, \&
  {Voronkov}}]{purser_mnras_460_2016}
{Purser}, S.~J.~D., {Lumsden}, S.~L., {Hoare}, M.~G., {et~al.} 2016, \mnras,
  460, 1039

\bibitem[{{Rafikov}(2005)}]{2005ApJ...621L..69R}
{Rafikov}, R.~R. 2005, \apjl, 621, L69

\bibitem[{{Rafikov}(2007)}]{rafikov_apj_662_2007}
{Rafikov}, R.~R. 2007, \apj, 662, 642

\bibitem[{{Reg{\'a}ly} \& {Vorobyov}(2017)}]{regaly_aa_601_2017}
{Reg{\'a}ly}, Z. \& {Vorobyov}, E. 2017, \aap, 601, A24

\bibitem[{{Reiter} {et~al.}(2017){Reiter}, {Kiminki}, {Smith}, \&
  {Bally}}]{reiter_mnras_470_2017}
{Reiter}, M., {Kiminki}, M.~M., {Smith}, N., \& {Bally}, J. 2017, \mnras, 470,
  4671

\bibitem[{{Rosdahl} \& {Teyssier}(2015)}]{rosdhal_mnras_449_2015}
{Rosdahl}, J. \& {Teyssier}, R. 2015, \mnras, 449, 4380

\bibitem[{{Rosen} {et~al.}(2016){Rosen}, {Krumholz}, {McKee}, \&
  {Klein}}]{rosen_mnras_463_2016}
{Rosen}, A.~L., {Krumholz}, M.~R., {McKee}, C.~F., \& {Klein}, R.~I. 2016,
  \mnras, 463, 2553

\bibitem[{{Samal} {et~al.}(2018){Samal}, {Chen}, {Takami}, {Jose}, \&
  {Froebrich}}]{samal_mnras_477_2018}
{Samal}, M.~R., {Chen}, W.~P., {Takami}, M., {Jose}, J., \& {Froebrich}, D.
  2018, \mnras, 477, 4577

\bibitem[{{Seifried} {et~al.}(2011){Seifried}, {Banerjee}, {Klessen}, {Duffin},
  \& {Pudritz}}]{seifried_mnras_417_2011}
{Seifried}, D., {Banerjee}, R., {Klessen}, R.~S., {Duffin}, D., \& {Pudritz},
  R.~E. 2011, \mnras, 417, 1054

\bibitem[{{Seifried} {et~al.}(2013){Seifried}, {Banerjee}, {Pudritz}, \&
  {Klessen}}]{seifried_mnras_432_2013}
{Seifried}, D., {Banerjee}, R., {Pudritz}, R.~E., \& {Klessen}, R.~S. 2013,
  \mnras, 432, 3320

\bibitem[{{Seifried} {et~al.}(2015){Seifried}, {Banerjee}, {Pudritz}, \&
  {Klessen}}]{seifried_mnras_446_2015}
{Seifried}, D., {Banerjee}, R., {Pudritz}, R.~E., \& {Klessen}, R.~S. 2015,
  \mnras, 446, 2776

\bibitem[{{Seifried} {et~al.}(2012){Seifried}, {Pudritz}, {Banerjee}, {Duffin},
  \& {Klessen}}]{seifried_mnras_422_2012}
{Seifried}, D., {Pudritz}, R.~E., {Banerjee}, R., {Duffin}, D., \& {Klessen},
  R.~S. 2012, \mnras, 422, 347

\bibitem[{{Stecklum} {et~al.}(2017){Stecklum}, {Heese}, {Wolf}, {Garatti},
  {Ibanez}, \& {Linz}}]{stecklum_2017a}
{Stecklum}, B., {Heese}, S., {Wolf}, S., {et~al.} 2017, ArXiv e-prints
  [\eprint[arXiv]{1712.01451}]

\bibitem[{{Teyssier}(2002)}]{teyssier_aa_385_2002}
{Teyssier}, R. 2002, \aap, 385, 337

\bibitem[{{Vaidya} {et~al.}(2011){Vaidya}, {Fendt}, {Beuther}, \&
  {Porth}}]{vaidya_apj_742_2011}
{Vaidya}, B., {Fendt}, C., {Beuther}, H., \& {Porth}, O. 2011, \apj, 742, 56

\bibitem[{{Vaidya} {et~al.}(2018){Vaidya}, {Mignone}, {Bodo}, {Rossi}, \&
  {Massaglia}}]{vaidya_apj_865_2018}
{Vaidya}, B., {Mignone}, A., {Bodo}, G., {Rossi}, P., \& {Massaglia}, S. 2018,
  \apj, 865, 144

\bibitem[{{Vorobyov}(2016)}]{vorobyov_aa_590_2016}
{Vorobyov}, E.~I. 2016, \aap, 590, A115

\bibitem[{{Vorobyov} \& {Basu}(2005)}]{vorobyov_apj_633_2005}
{Vorobyov}, E.~I. \& {Basu}, S. 2005, \apjl, 633, L137

\bibitem[{{Vorobyov} \& {Basu}(2006)}]{voroboyov_apj_650_2006}
{Vorobyov}, E.~I. \& {Basu}, S. 2006, \apj, 650, 956

\bibitem[{{Vorobyov} \& {Basu}(2010)}]{vorobyov_apj_719_2010}
{Vorobyov}, E.~I. \& {Basu}, S. 2010, \apj, 719, 1896

\bibitem[{{Vorobyov} \& {Basu}(2015)}]{vorobyov_apj_805_2015}
{Vorobyov}, E.~I. \& {Basu}, S. 2015, \apj, 805, 115

\bibitem[{{Vorobyov} {et~al.}(2013){Vorobyov}, {DeSouza}, \&
  {Basu}}]{vorobyov_apj_768_2013}
{Vorobyov}, E.~I., {DeSouza}, A.~L., \& {Basu}, S. 2013, \apj, 768, 131

\bibitem[{{Vorobyov} \& {Elbakyan}(2018)}]{2018A&A...618A...7V}
{Vorobyov}, E.~I. \& {Elbakyan}, V.~G. 2018, \aap, 618, A7

\end{thebibliography}


\end{document}